\documentclass[onecolumn,superscriptaddress,preprintnumbers,nofootinbib,amsmath,noshowpacs,aps,pra,floatfix,longbibliography]{revtex4-2}
\usepackage[utf8]{inputenc}
\usepackage{graphicx}
\usepackage{dcolumn}
\usepackage{bm}
\usepackage{float}
\usepackage{amssymb}
\usepackage{mathtools}
\usepackage{array}
\usepackage{fixmath}
\usepackage{subcaption}
\usepackage{mathrsfs}
\usepackage{xcolor}
\usepackage{bbm}
\usepackage{physics}
\usepackage{multirow}
\usepackage{upgreek}
\usepackage{mathdots}
\usepackage{appendix}
\usepackage{verbatim}
\usepackage[pdftex,colorlinks,linkcolor=black,citecolor=magenta,urlcolor=black]{hyperref}
\usepackage{cleveref}
\usepackage[export]{adjustbox}

\newcommand{\prep}{\textsc{Prep}}
\newcommand{\sel}{\textsc{Sel}}

\begin{document}

\title{Fault-tolerant quantum algorithms for simulating atomic nuclei}

\author{James Benstead}
\affiliation{AWE Nuclear Security Technologies, Aldermaston, RG7 4PR, UK}
\affiliation{School of Mathematics and Physics, University of Surrey, Guildford, Surrey GU2 7XH, UK}

\author{Michael Garn}
\affiliation{The Hartree Centre, STFC, Sci-Tech Daresbury, Warrington, WA4 4AD, UK}

\author{Neil Gaspar}
\affiliation{AWE Nuclear Security Technologies, Aldermaston, RG7 4PR, UK}

\author{Sean Greenaway}
\affiliation{PsiQuantum, Daresbury, WA4 4FS, UK}
\affiliation{PsiQuantum, 700 Hansen Way, Palo Alto, California, 94304, USA}

\author{Angus Kan}
\affiliation{PsiQuantum, Daresbury, WA4 4FS, UK}
\affiliation{PsiQuantum, 700 Hansen Way, Palo Alto, California, 94304, USA}

\author{Lloyd La Ronde}
\affiliation{School of Mathematics and Physics, University of Surrey, Guildford, Surrey GU2 7XH, UK}

\author{Chandan Sarma}
\affiliation{School of Mathematics and Physics, University of Surrey, Guildford, Surrey GU2 7XH, UK}

\author{Paul Stevenson}
\affiliation{School of Mathematics and Physics, University of Surrey, Guildford, Surrey GU2 7XH, UK}
\affiliation{AWE Nuclear Security Technologies, Aldermaston, RG7 4PR, UK}

\date{\today}

\begin{abstract}
To maximize the value of fault-tolerant quantum computers, it is essential to develop concrete applications beyond well-established domains such as chemistry and condensed-matter physics. Here we construct and compile quantum algorithms to simulate the structure of atomic nuclei -- a topic that has received relatively little attention from the quantum computing community despite its similarities to the electronic structure problem in chemistry -- via effective shell-model Hamiltonians and no-core-shell-model Hamiltonians with three-body interactions derived from chiral effective field theory. Furthermore, we provide quantum resource estimates, in terms of Toffoli gate and qubit counts, for these algorithms, which, to our knowledge, are the first such estimates for fault-tolerant quantum simulation of atomic nuclei. Notably, the estimates for $^{32}$Mg and $^{219}$At shell-model Hamiltonians are comparable to recent estimates of Femoco simulations, a standard benchmark in chemistry. For no-core-shell-model Hamiltonians suitable for light nuclei (up to $^{40}$Ca or so), we find that resource requirements are significantly higher, suggesting that more bespoke strategies are required to make such simulations practicable. Throughout this work, we draw upon the similarities between nuclear and electronic structure problems, while also highlighting challenges that are specific to the former. We hope this work will spur long-term collaborations between the nuclear and quantum computing community with the ultimate goal of realizing useful nuclear simulations on quantum computers.
\end{abstract}

\maketitle

%\tableofcontents

\section{Introduction}

The advantages and methods of simulating fermionic systems on fault-tolerant quantum computers are well studied, particularly in the context of the electronic structure problem from chemistry and Hubbard models from condensed matter physics~\cite{doi:10.1073/pnas.1619152114,PhysRevX.8.041015,Berry2019qubitizationof,Kivlichan2020improvedfault,Motta_2021,PhysRevResearch.3.033055,Campbell_2022,doi:10.1021/acs.jctc.4c00352,Kan_2025,PRXQuantum.2.030305,pb2g-j9cw,yngp-5fpm}. However, the study of fermions extend beyond the study of interacting electrons in molecules or solid-state systems. Here we consider the nuclear structure problem, which is the study of interacting nucleons, i.e., protons and neutrons, which are also fermions, in atomic nuclei.

The origins of the nuclear many-body problem date back to the identification of the nuclear atomic model by Rutherford, Geiger and Marsden~\cite{gegier_diffuse_1909,rutherford_scattering_1911} and the subsequent discovery of the neutron~\cite{chadwick_existence_1932}. This was followed by a great combination of experimental and theoretical work, which by the end of the 1930s had given rise to the concept of isospin~\cite{heisenberg_uber_1932,wigner_consequences_1937}, the semi-empirical mass formula~\cite{weizsacker_zur_1935,bethe_stationary_1936}, the compound nuclear model~\cite{bohr_neutron_1936}, and first theories of the nuclear force~\cite{majorana_uber_1933,yukawa,fermi_tentativo_1934}.  

Continued development of the microscopic understanding of nuclei led to unifying ideas such as the use of symmetries and the nuclear shell model, which were recognized with the 1963 Nobel Prize in Physics~\cite{wigner_consequences_1937,mayer_jensen}. Over time, the reach of theoretical physics extended beyond the energy scale characteristic of nuclear structure, i.e., MeV, into the realm of particle physics, i.e., GeV. Yet, many open questions in nuclear structure remain, such as those concerning the limits of nuclear stability. For example, although it is known that atomic nuclei with a ``magic" number of protons or neutrons are much more stable than others, it remains difficult to a priori predict how many nucleons can be added to these stable nuclei before they become unbound~\cite{heyde1990nuclear}. The mechanisms that give rise to different nuclear shapes, e.g., spherical and deformed configurations, are also not well understood. Addressing these questions is crucial, as insights from nuclear structure provide critical input to related fields including nuclear astrophysics, nuclear reactions, and the manifold industrial applications of nuclear data~\cite{APRAHAMIAN2005535,goriely2023nuclear,THIELEMANN2003139,goriely2008improved,jenkins2025}.

One way to improve our understanding of nuclear structure is to numerically solve a many-body Sch\"odinger equation for atomic nuclei. Due to the similarities between electronic and nucleonic problems, at a high level, similar classical methodologies, e.g., configuration-interaction  formalism~\cite{whitehead_numerical_1972,shimizu_thick-restart_2019}, coupled-cluster~\cite{PhysRevC.69.054320,PhysRevC.89.014319,Hagen_2014}, density-functional theory~\cite{colo2020nuclear}, Monte Carlo~\cite{RevModPhys.87.1067} and density matrix renormalization group (DMRG)/matrix product state (MPS)~\cite{PhysRevLett.97.110603,PhysRevC.79.014304,TICHAI2023138139,TICHAI2024138841,Dukelsky_2004,PhysRevC.78.041303,PhysRevC.92.051303,PhysRevC.65.054319,gibbs_low_2026}, are often employed to solve them. At a lower level, these algorithms, when applied to nuclear structure, will have to incorporate some specific characteristics that differentiate the problem from electronic systems. First, in nuclei, there is no ionic background, but there are two types of fermions -- protons and neutrons -- distinguishable from each other. Second, nuclei are self-bound systems in which the combination of the strong nuclear interaction and the Pauli principle conspire to give rather weakly-bound systems (e.g., the simplest nucleus, the deuteron, which consists of one proton and one neutron, does not have any bound excited states). Furthermore, the nuclear interaction is not as well-understood as the Coulomb interaction between electrons is. In part, this is a result of the underlying theory, i.e., quantum chromodynamics (QCD), being challenging, and in part because the force between composite nucleons is a kind of residual interaction ``leaking out'' of each nucleon in which quarks are strongly confined. Evidence for a physical three-body force in nuclei is strong, while leading theories for deriving forces from underlying QCD ideas, e.g., chiral effective field theory~\cite{weinberg_1979,weinberg_1990,epelbaum2020high,Machleidt_2011,RevModPhys.81.1773}, also suggest higher-order forces are present. Such forces manifest as $n(>2)$-body terms in second-quantized nuclear structure Hamiltonians~\cite{HEBELER20211,epelbaum2020high,Machleidt_2011,RevModPhys.81.1773,Miyagi_2023}. Furthermore, the exponential scaling of Hilbert space with particle number is a difficulty in nuclear structure as it is in any many-body system. These various considerations make atomic nuclei a challenging system for calculation or simulation by any method.

Despite the challenges in and the importance of the nuclear structure problem, as well as its similarity to electronic problems, it has received much less attention from the quantum computing community. While there are some papers that study near-term quantum heuristic designed to run on noisy quantum devices~\cite{PhysRevLett.120.210501,perez2023nuclear,Bhoy_2024,PhysRevC.108.064305,gu2025scalablequantumcomputationsatomic,sarma2025low,maheshwari2025single,44k6-w3dt,siwach_quantum_2021,siwach_quantum_2022,singh_quantum_2024,costa_quantum_2025}, only a small amount of work on quantum algorithms for fault-tolerant quantum computers exists ~\cite{gibbs_low_2026}~\footnote{There are, however, fault-tolerant algorithms that target other models relevant to nuclear physics but are not tailored to study nuclear structure~\cite{watson2025quantum,spagnoli2025quantum,liu2025efficient}.}. In contrast, the chemistry and condensed-matter community have largely shifted efforts to fault-tolerant algorithms, embracing a compelling view that fault-tolerant quantum computers are the only scalable quantum computational approach that we know of. As a result of such efforts, fault-tolerant simulations of molecules and condensed-matter systems have become orders of magnitude cheaper over the last decade~\cite{doi:10.1073/pnas.1619152114,PhysRevX.8.041015,Berry2019qubitizationof,Kivlichan2020improvedfault,Motta_2021,PhysRevResearch.3.033055,Campbell_2022,Kan_2025,PRXQuantum.2.030305,pb2g-j9cw,yngp-5fpm}. Now, the chemistry and condensed-matter community are poised to be the first disciplines to benefit from the eventual arrival of fault-tolerant quantum computers.

In this work, we develop new tools to study nuclear structure via fault-tolerant quantum algorithms. In particular, our main contributions are as follows. We construct qubitization-based energy estimation algorithms~\cite{PhysRevLett.121.010501} for nuclear structure problems. These algorithms can, for example though not exclusively, be used to perform ground state energy estimation -- an application where quantum computers are expected to offer significant advantage over classical computers for fermionic systems in chemistry and condensed matter~\cite{Lee_2023}, and we expect, in nuclear physics as well. We further compile these algorithms and provide quantum resource estimates in terms of Toffoli gate and qubit counts, which are cost metrics relevant to fault tolerant quantum computers; to our knowledge, these are the first known quantum resource estimates for nuclear structure problems. Furthermore, our circuit constructions improve significantly over prior art~\cite{liu2025blockencodinglowgate}.

The rest of the paper is organized as follows. In section~\ref{sec:methods}, we provide an overview of our choice of nuclear Hamiltonians and the qubitization-based phase estimation algorithm. We present our circuit implementation of the algorithm and quantum resource estimates in section~\ref{sec:results}. Finally, we conclude and discuss our results and future directions in section~\ref{sec:discussion}.

\section{Methods}
\label{sec:methods}

\subsection{Target Hamiltonians}
\label{sec:Ham}

Over a vast range of energy, the structure of atomic nuclei can be effectively studied using a non-relativistic Hamiltonian with nucleons, i.e., protons and neutrons, as the only degrees of freedom. In particular, the Hamiltonian of an atomic nucleus, which we hereafter refer to as the nuclear Hamiltonian, generally takes the form
\begin{equation}\label{eq:NuHamil}
    H = \sum_{p,q} t_{pq} a^\dagger_p a_q + \sum_{p<q, r<s} V_{pqrs} a^\dagger_p a^\dagger_q a_r a_s + \sum_{p<q<r, s<t<u} W_{pqrstu} a^\dagger_p a^\dagger_q a^\dagger_r a_s a_t a_u.
\end{equation}
Here, the physics of each term depends on the version of the shell model, with more details given below, but note that the nuclear Hamiltonian can comprise not only one and two-body terms, like the electronic structure Hamiltonian, but also a three-body term, which is absent in the electronic structure Hamiltonian.

There are different types of Hamiltonian used in shell model calculation: empirical interactions with matrix elements fitted to experimental data~\cite{wildenthal_empirical_1984};  those derived using phenomenological interactions~\cite{PhysRevC.63.024001,PhysRevC.51.38,PhysRevC.79.014310,PhysRevC.43.602,PhysRevC.74.034315}, and those derived from underlying theory, nowadays unified under the chiral effective field theory (EFT) framework~\cite{RevModPhys.81.1773,Machleidt_2011,HEBELER20211,Miyagi_2023,epelbaum2020high}. Below, we give brief descriptions of them; for the inquisitive readers, please refer to~\cite{epelbaum2020high,Machleidt_2011,RevModPhys.81.1773,RevModPhys.87.1067,Miyagi_2023} and references therein. 

Empirical interactions are numerically fitted to various nuclear data - typically a subset of output observables desired.  Such fits take place at the level of interaction matrix elements.  Phenomenological potential-based interactions abstract the problem to parameterization in terms of physics-inspired effective interactions.  The empirical and phenomenological Hamiltonians typically do not contain a three-body term. Furthermore, in the shell model~\cite{mayer_jensen} or configuration interaction context, they are typically built from an inert core, which is usually ``doubly magic", i.e., contains a magic number of protons and neutrons. The inert core is analogous to the classical nucleus for the electronic structure problem under the Born-Oppenheimer approximation. We will consider phenomenological Hamiltonians for $^{24}$Mg~\cite{PhysRevC.74.034315}, $^{32}$Mg~\cite{PhysRevC.79.014310}, $^{219}$At~\cite{PhysRevC.43.602}. For the Mg nuclei, an inert $^{16}$O core is assumed, resulting in an 8-particle, 24-orbital Hamiltonian for $^{24}$Mg, and a 16-particle, 64-orbital Hamiltonian for $^{32}$Mg. For $^{219}$At, we assume $^{208}$Pb as the core, leading to a 13-particle, 102-orbital Hamiltonian. We note that even though $^{24}$Mg is a classically solvable problem, its resource requirements set a reasonable expectation for the minimal specifications of a FTQC that can perform meaningful nuclear calculations that can be crosschecked with classical results. $^{32}$Mg is chosen as a challenging nucleus with relatively low mass. It is sometimes used as a limiting case for classical solvers where configuration truncation is needed~\cite{shimizu_thick-restart_2019}.  $^{219}$At is chosen as a classically-challenging problem in a large model space, but for which an effective interaction has been developed. 

EFT Hamiltonians, unlike phenomenological interactions, typically do not assume an inert core, i.e., are no-core shell models, and are built ground-up from interaction terms that naturally arise in the considered EFT. In particular, Chiral EFT~\cite{weinberg_1979,weinberg_1990}, which is often employed to study the low-energy phenomena of QCD such as hadrons and thus nucleons, enables a systemic derivation of multi-nucleon interactions via a scheme known as \emph{power-counting}, which generated an understanding of why three-body interactions are a small fraction ($\sim 10\%)$ of two-body interactions and interactions involving $n>3$ bodies are negligible in most nuclear scenarios and tend to be too computationally intensive to derive~\cite{Miyagi_2023,epelbaum2020high,HEBELER20211,Machleidt_2011,RevModPhys.81.1773}. Note that in the derivations of Chiral EFT interactions, a set of low-energy constants are fitted to nucleonic scattering data. 

We generate our choices of Chiral EFT Hamiltonians, using the \texttt{NuHamil} software~\cite{Miyagi_2023}, where the two and three-body interactions are from~\cite{PhysRevC.68.041001,PhysRevC.101.014318} and a similarity renormalization group ramp space from~\cite{PhysRevC.90.024325}. Furthermore, we set the \texttt{NuHamil} energy truncation parameters $e_{\max}=8$ and $e_{2\max}=16$, and $e_{3\max} = 8$~\footnote{Larger $e_{3\max}$ values can lead to massive interaction matrices (hundreds of Gb of data)~\cite{PhysRevC.90.024325}, which become demanding to process classically. So, while they can in principle be accommodated by our algorithms, we stick to smaller $e_{3\max}$ values in this work.}, and set the resolution scale for similarity renormalization group evolution to be $\lambda = 2 \text{ fm}^{-1}$. It is worth pointing out that for a given nucleus, no-core shell-model Hamiltonians, such as EFT Hamiltonians, because of their ground-up nature, typically comprise more orbitals and thus are much denser than conventional shell models that contain a core. Therefore, in practice, no-core shell models are more often used for lighter nuclei; our choices of Hamiltonians are sufficient for up to $^{40}$Ca or so~\cite{Miyagi_2023}. For our choice of model space truncation, we have 660 orbitals (330 for protons, 330 for neutrons)- a size considerably larger than the effective interaction problems we are considering above, and in which classical calculations required drastic approximation methods, e.g. limiting configuration via some heuristic.  As an indication of the problem size, the number of configurations possible for 20 protons in 330 orbitals and 20 neutrons in 330 orbitals is $\left(330\atop20\right)^2\simeq2^{211}$.

%$e_{3\max} = 8$, the number of orbitals is $\ins{XX}$ and for $e_{3\max}=10$, the number of orbitals is \ins{XX}. \ins{justification for why these Hamiltonians are chosen.}

We stress that in this work, we simply employ instances of nuclear Hamiltonians for the purpose of generating quantum resource estimates for our algorithms, and will not delve into the properties of such Hamiltonians from a nuclear-physics perspective. For more details on our choice of Hamiltonians, please read appendix~\ref{app:nuHam}.

\subsection{Qubitization-based energy estimation algorithm}
\label{sec:qpe}

Here we review a qubitization-based quantum phase estimation algorithm that can be used to estimate the energy of an input state with respect to a Hamiltonian~\cite{Low2019hamiltonian,PhysRevLett.121.010501,PhysRevX.8.041015}. Qubitization~\cite{Low2019hamiltonian} assumes that the Hamiltonian $H$ is encoded as a linear combination of unitaries (LCU), using the following oracles:
\begin{gather}
    \prep \ket{0}_a = \sum_{i=1}^L \sqrt{\frac{c_i}{\lambda}} \ket{i}_a, \: \lambda = \sum_{i=1}^L |c_i|, \\
    \sel = \sum_{i=1}^L \ketbra{i}{i}_a\otimes U_i, \\
    \text{where } (\bra{0}_a\otimes \bra{0}_b \otimes I_s)\prep^\dagger \cdot \sel \cdot \prep (\ket{0}_a\otimes \ket{0}_b \otimes I_s) = \sum_{i=1}^L \frac{c_i}{\lambda} O_i \equiv \frac{H}{\lambda},
\end{gather}
where $\lambda$ is sometimes referred to as the LCU 1-norm. Furthermore, $U_i$ is a unitary, which is implemented as a quantum circuit that acts on registers $b,s$ where $s$ is the system register, on which $O_i$ and thus $H$ act. $H$ is then encoded in the $\ket{0}_a\otimes \ket{0}_b$ ancilla subspace of $\prep^\dagger \cdot \sel \cdot \prep$, which is commonly known as a $\emph{block-encoding}$ of $H$. Note that sometimes when $O_i$ is unitary, it can be preferable to directly implemented it as a unitary circuit, then $U_i = O_i$ and the ancilla register $b$ will no longer be necessary.

Using $\prep$ and $\sel$, one can construct a quantum phase estimation algorithm~\cite{PhysRevLett.121.010501} that underpins many modern quantum algorithms designed to solve fermionic models~\cite{PhysRevX.8.041015,Berry2019qubitizationof,PhysRevResearch.3.033055,Kan_2025,PRXQuantum.2.030305,pb2g-j9cw,yngp-5fpm}. The algorithm queries a walk operator $\mathcal{W}$, which shares the same eigenvalues as $e^{i \arccos{(H/\lambda)}}$ and $e^{-i \arccos{(H/\lambda)}}$ and can be constructed as
\begin{equation}\label{circ:walk}
    \includegraphics[height=3.5cm,valign=c]{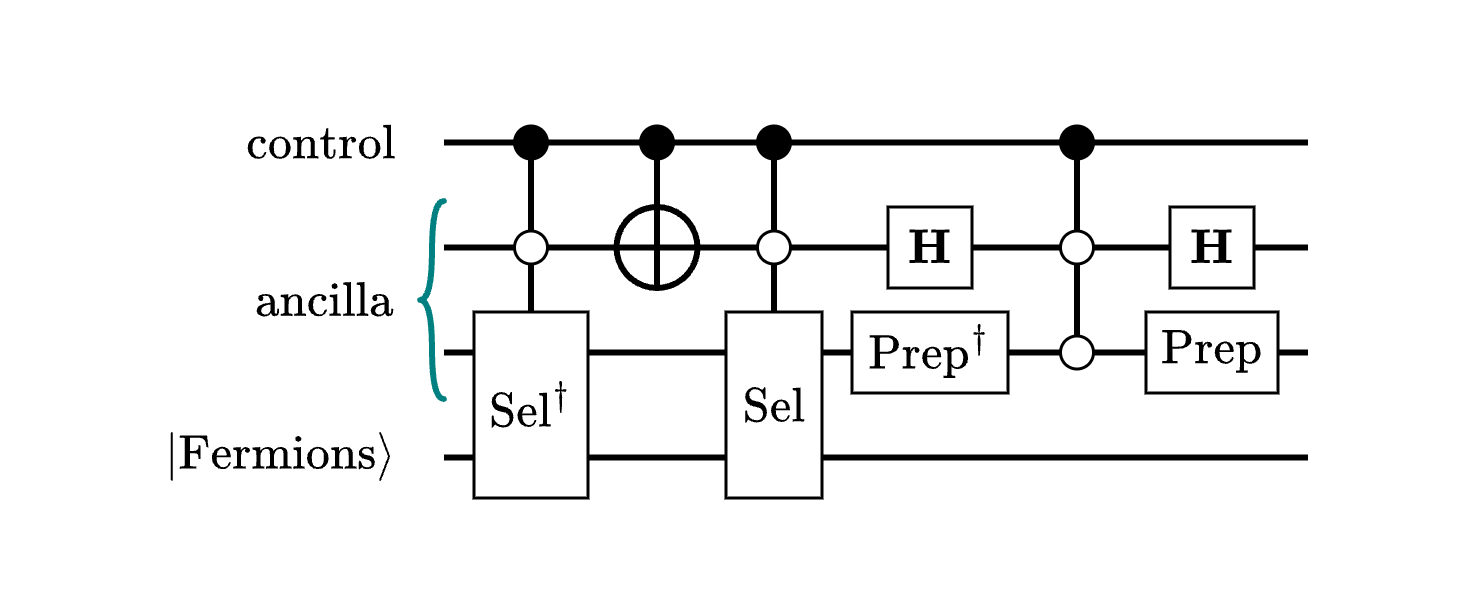},
\end{equation}
where the first three operations can be interpreted as a self-inverse $\sel$ for $\frac{1}{2} (H^\dagger + H) = H$ and the remaining operations form another self-inverse unitary. If $\sel = \sel^\dagger$, the first two operations and the top ancilla qubit can be removed. The total cost of such a quantum phase estimation algorithm can then be estimated as
\begin{equation}
    C = Q_\mathcal{W} \cdot C_\mathcal{W},
\end{equation}
where $Q_\mathcal{W}$ and $C_\mathcal{W}$ are the query count of $\mathcal{W}$ and cost per $\mathcal{W}$. The costs of other parts of the algorithm, e.g., quantum Fourier transform, are negligible in comparison and hence, typically not estimated~\cite{PhysRevX.8.041015,Berry2019qubitizationof,PhysRevResearch.3.033055,Kan_2025,PRXQuantum.2.030305,pb2g-j9cw,yngp-5fpm}. We review how to compute $Q_\mathcal{W}$ in appendix~\ref{app:query}, and discuss the circuit construction of $\mathcal{W}$ and derivation of $C_\mathcal{W}$ in section~\ref{sec:circuits}.

\section{Results}
\label{sec:results}

\subsection{Quantum circuits}
\label{sec:circuits}

In this section, we construct the $\sel$ and $\prep$ quantum circuits that we use to simulate the nuclear Hamiltonian. We begin with the Jordan-Wigner transformation~\cite{jw}:
\begin{equation}\label{eq:JW}
    a^\dagger_p \mapsto \sigma^+_p \prod_{p>i}Z_i, \: a_p \mapsto \sigma^-_p\prod_{p>i}Z_i,
\end{equation}
where $\sigma^- = \ketbra{0}{1}$ and $\sigma^+ = \ketbra{1}{0}$, and the fermions are represented in the occupation basis:
\begin{equation}\label{eq:occ_b}
    \ket{\text{fermions}} = \ket{f_1}\ket{f_2}...\ket{f_{N}} = \bigotimes_{i=1}^{N} \ket{f_i},
\end{equation}
where $N$ is the number of orbitals, and $f_i = 0$ or $1$ if the $i$th orbital is empty or occupied respectively. Then, pairs of fermionic operators can be mapped to spin-1/2 operators as follows:
\begin{align}\label{eq:JW_pm}
    a^\dagger_p a_q &\mapsto 
    \begin{cases}
        (Z_q \ketbra{0}{1}) Z_{q+1}...Z_{p-1} (\ketbra{1}{0})_p \text{ if }p>q \\
        (\ketbra{1}{0}Z_p) Z_{p+1}...Z_{q-1} (\ketbra{0}{1})_q \text{ if }q>p\\
        \ketbra{1}{1} \text{ if }p=q
    \end{cases} \nonumber \\
    &= 
    \begin{cases}
        (\ketbra{0}{1})_q Z_{q+1}...Z_{p-1} (\ketbra{1}{0})_p \text{ if }p>q \\
        (\ketbra{1}{0})_p Z_{p+1}...Z_{q-1} (\ketbra{0}{1})_q \text{ if }q>p\\
        \ketbra{1}{1} \text{ if }p=q
    \end{cases},
\end{align}
where the string of $Z$ vanishes if $|p-q|=1$. Similarly, one can derive
\begin{align}\label{eq:JW_pp}
    a^\dagger_p a^\dagger_q &\mapsto 
    \begin{cases}
        (Z_q \ketbra{1}{0}) Z_{q+1}...Z_{p-1} (\ketbra{1}{0})_p \text{ if }p>q \\
        (\ketbra{1}{0}Z_p) Z_{p+1}...Z_{q-1} (\ketbra{1}{0})_q \text{ if }q>p
    \end{cases} \nonumber \\
    &= 
    \begin{cases}
        -(\ketbra{1}{0})_q Z_{q+1}...Z_{p-1} (\ketbra{1}{0})_p \text{ if }p>q \\
        (\ketbra{1}{0})_p Z_{p+1}...Z_{q-1} (\ketbra{1}{0})_q \text{ if }q>p
    \end{cases}.
\end{align}
And similarly,
\begin{align}\label{eq:JW_mm}
    a_p a_q &\mapsto 
    \begin{cases}
        (\ketbra{0}{1})_q Z_{q+1}...Z_{p-1} (\ketbra{0}{1})_p \text{ if }p>q \\
        -(\ketbra{0}{1})_p Z_{p+1}...Z_{q-1} (\ketbra{0}{1})_q \text{ if }q>p
    \end{cases}.
\end{align}
As a result, the transformed Hamiltonian can be written as
\begin{align}\label{eq:jw_ham}
    H&=\sum_{p,q} t_{p,q} (\sigma^+_p \vec{Z}_{pq}\sigma^-_q)  + \sum_{p<q, r<s} V_{pqrs} (\sigma^+_p \vec{Z}_{pq}\sigma^+_q)(\sigma^-_r \vec{Z}_{rs}\sigma^-_s) \nonumber \\
    &\quad + \sum_{p<q<r, s<t<u} W_{pqrstu} (\sigma^+_p \vec{Z}_{pq}\sigma^+_q)(\sigma^+_r \vec{Z}_{rs}\sigma^-_s)(\sigma^-_t \vec{Z}_{tu}\sigma^-_u)
\end{align}
where $\vec{Z}_{pq}$ denotes any string of $Z$ operators between fermions $p$ and $q$, including any overall minus sign, due to the Jordan-Wigner transformation.

Let us first address the $\sel$ operation. Our $\sel$ for the one-body term $a^\dagger_p a_q$ is a product of two circuits: $O^{+-}_f O^{+-}_Z$, where $O^{+-}_Z$ is given by
\begin{widetext}
\begin{equation}\label{circ:ozpm}
    \includegraphics[height=5.5cm,valign=c]{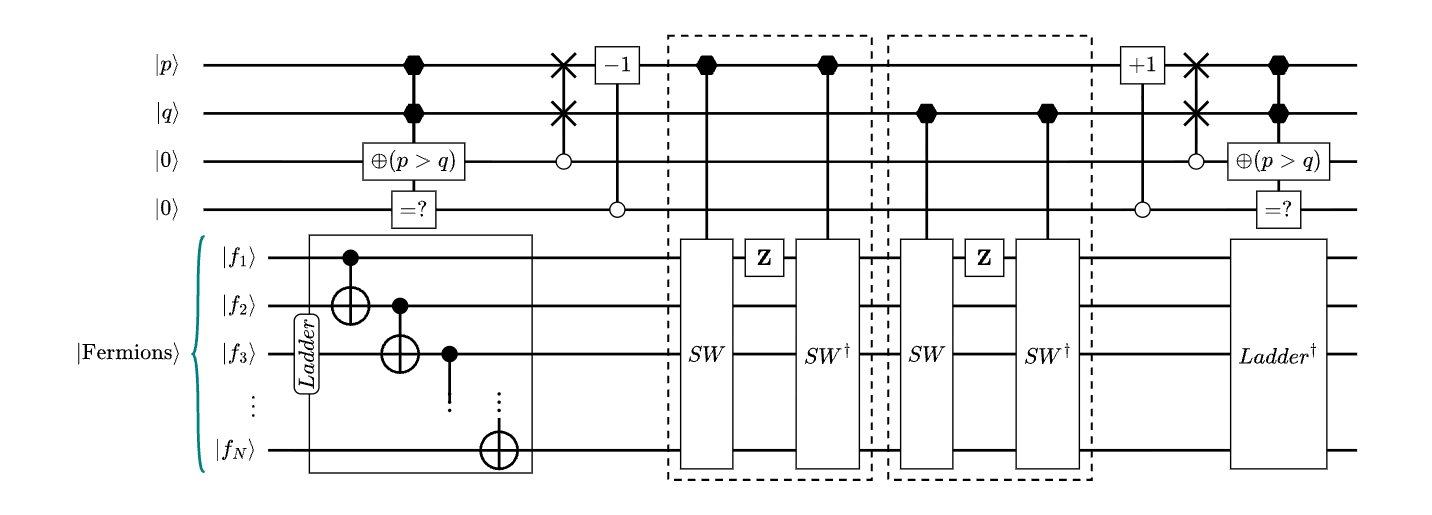},
\end{equation}
\end{widetext}
and $O^{+-}_f$ is given by
\begin{equation}\label{circ:ofpm}
    \includegraphics[height=4.5cm,valign=c]{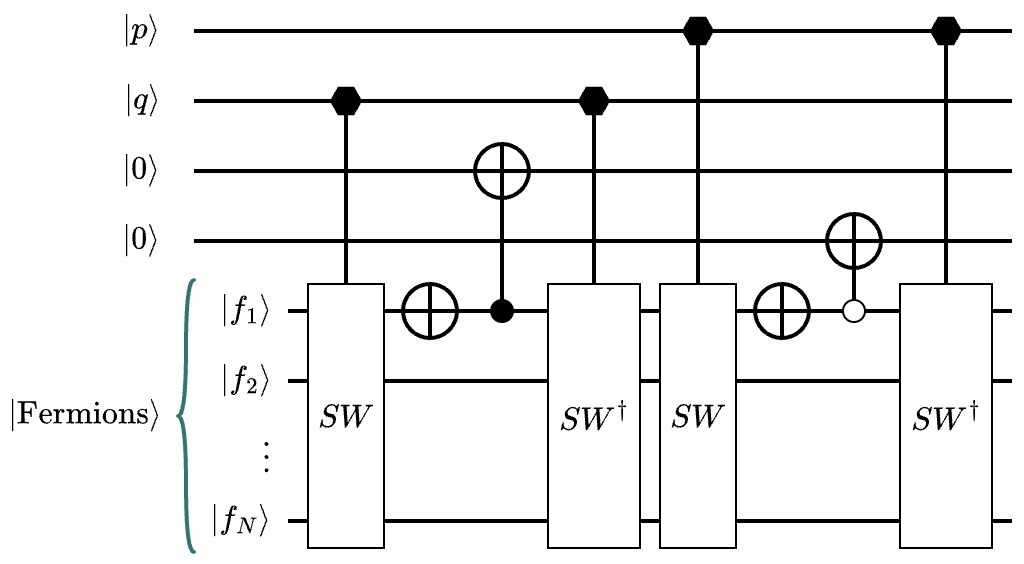},
\end{equation}
where we hereafter use the hexagonal controls to indicate multiplexing variables or control registers with more than one qubit, and $SW$ denotes a SwapUp operation~\cite{Low2024tradingtgatesdirty}:
\begin{equation}
    \text{SwapUp} \ket{p} \ket{f_1}\ket{f_2}...\ket{f_N} = \ket{p} \ket{f_p} \ket{\ast}...\ket{\ast}.
\end{equation}
In terms of gate cost, the SwapUp operations are the dominant part, as they cost $N$ Toffoli gates each~\cite{Low2024tradingtgatesdirty}, whereas the register-swaps, adders and comparators cost $\log_2(N) + O(1)$~\cite{Gidney2018halvingcostof}. In $O_Z^{+-}$, prior to the SwapUps, the ancilla index registers $\ket{p}$ and $\ket{q}$ are swapped so that the larger index is on the top. Furthermore, if $p\neq q$, the larger index is incremented so that the registers are:
\begin{equation}
    \begin{cases}
        \ket{p-1}\ket{q} \text{ if }p>q,\\
        \ket{q-1}\ket{p} \text{ if }q>p, \\
        \ket{p}\ket{q} \text{ if }p=q.
    \end{cases}
\end{equation}
The gates in the dashed boxes conjugated by the CNOTs are responsible for the $Z$ phasing: First, the $Ladder$ of CNOTs convert $\ket{\text{Fermions}}$ from the occupation basis to a parity basis, i.e., $\bigotimes_{i=1}^N \ket{f_i} \mapsto \bigotimes_{i=1}^N \ket{\oplus_{j=1}^i f_j}$. Then, the $Z$ gates conjugated by $SW$ and its inverse will apply
\begin{equation}
    \begin{cases}
        (\prod_{i=1}^{p-1}Z_i)(\prod_{i=1}^{q}Z_i) = \prod_{i=q+1}^{p-1}Z_i \text{ if }p>q, \\
        (\prod_{i=1}^{q-1}Z_i)(\prod_{i=1}^{p}Z_i) = \prod_{i=p+1}^{q-1}Z_i \text{ if }q>p, \\
        (\prod_{i=1}^{q}Z_i)(\prod_{i=1}^{p}Z_i) = 1 \text{ if }q=p,
    \end{cases}
\end{equation}
matching the phases in~\eqref{eq:JW_pm} as desired. The remaining operations in the circuit restore the ancilla registers to $\ket{0}$. In $O_f^{+-}$, $\ket{f_q}$ is first flipped, but an ancilla will be flagged if the flipped state $\ket{f_q\oplus 1} = \ket{1}$, meaning $\ket{f_q}=\ket{0}$. Similarly, an ancilla will be flagged if $\ket{f_p\oplus 1} = \ket{0}$. In other words, if the two flags are $\ket{00}$, $\ket{f_q}$ is first lowered and then $\ket{f_p}$ is raised as desired; any other operations on $\ket{f_q}$ and $\ket{f_p}$ are unphysical and are not in the block-encoded subspace, i.e., $\ket{00}$ ancilla space. 

While our circuits share some similarities with those in~\cite{Wan2021exponentiallyfaster,liu2025blockencodinglowgate}, they are more optimized. It was first shown in~\cite{Wan2021exponentiallyfaster} that the Jordan-Wigner $Z$-strings can be addressed using CNOT ladders and SwapUps, as we do in $O^{+-}_Z$. The key difference between our circuits is that we encode the non-unitary $\sigma^\pm$ operators directly in $O^{+-}_f$ without expanding them into the unitary Pauli operators, as done in~\cite{Wan2021exponentiallyfaster}. We have chosen to do so because we want to avoid encoding the increasingly large number of different Pauli strings that will inevitably arise from expanding monomials of fermionic operators in the two and especially, three-body term. The $\sel$ circuits in~\cite{liu2025blockencodinglowgate}, similar to ours, encode $\sigma^\pm$ directly, and consist of phasing and flipping parts. However, while we encode the $p\neq q$ and $p=q$ in one circuit, these two cases are treated separately and then combined as a LCU in~\cite{liu2025blockencodinglowgate}. Neglecting logarithmically sized components, the $\sel$ circuit from~\cite{liu2025blockencodinglowgate} for the $p\neq q$ case contains a phasing and flipping circuit analogous to $O_{Z}^{+-}$ and $O_{f}^{+-}$, respectively, while that for the $p=q$ case contains only $O_{f}^{+-}$ because the phase is trivial. For the $p\neq q$ case, their phasing and flipping circuits consist of 6 and 8 SwapUps, respectively, and the flipping circuit consists of 2 SwapUps for the $p=q$ case. Thus, overall our $\sel$ circuit for the one-body term achieves $\sim 50\%$ reduction in Toffoli count.

In the considered phenomenological shell-model Hamiltonians, by definition, the one-body term is diagonal:
\begin{equation}\label{eq:diag1bd}
    \sum_{p} t_p a^\dagger_p a_p \mapsto \sum_{p}t_p \frac{1-Z_p}{2}.
\end{equation}
In this case, the $O^{+-}_f O^{+-}_Z$ circuit is an overkill. We can subtract a constant factor of $\frac{1}{2}$ and implement the $\sel$ as a QROM~\cite{PhysRevX.8.041015}: $\sum_p \ketbra{p}{p} \otimes Z_p$, which costs only $N+O(1)$ Toffoli gates. Another added benefit is that the 1-norm of the one-body term is then halved, i.e., from $\sum_p |t_p|$ to $\sum_p \frac{|t_p|}{2}$.

The two-body $\sel$ circuit is a product of two smaller circuits, which are similar to the one-body $\sel$ circuit. The first circuit is $O^{--}_f O^{--}_Z$, where $O^{--}_Z$ is
\begin{equation}\label{circ:ozmm}
    \includegraphics[height=4.5cm,valign=c]{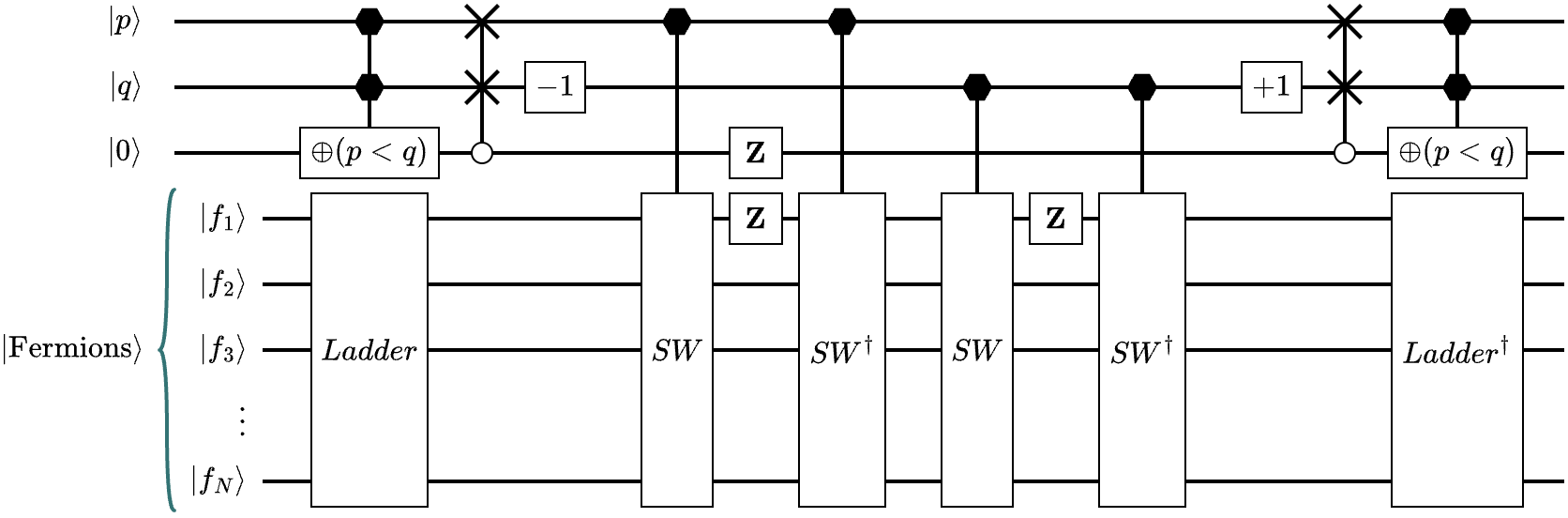},
\end{equation}    
and $O^{--}_f$ is
\begin{equation}\label{circ:ofmm}
    \includegraphics[height=4.5cm,valign=c]{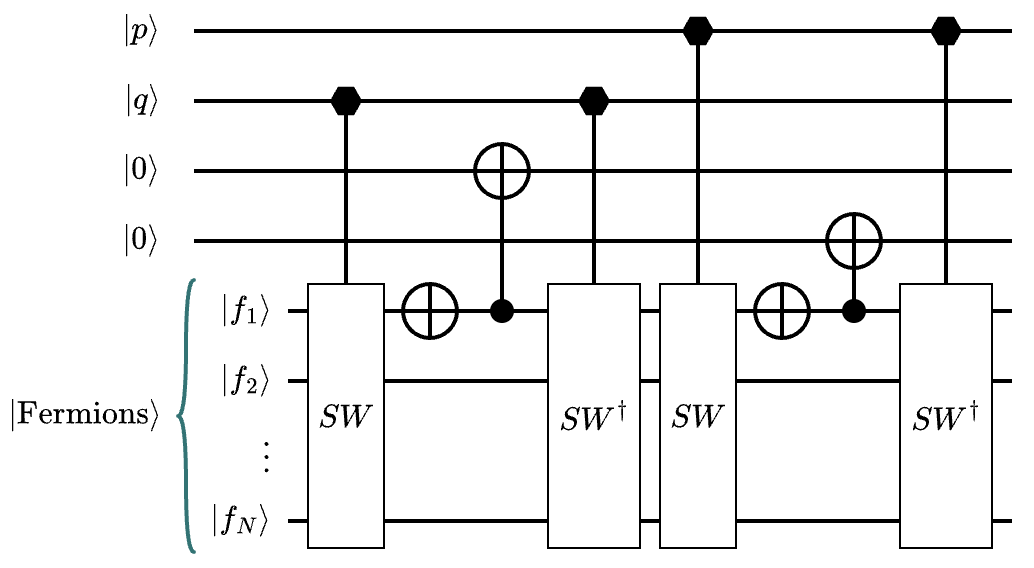},
\end{equation} 
and the second circuit is $O^{++}_f O^{++}_Z$, where $O_Z^{++}$ is
\begin{equation}\label{circ:ozpp}
    \includegraphics[height=5.5cm,valign=c]{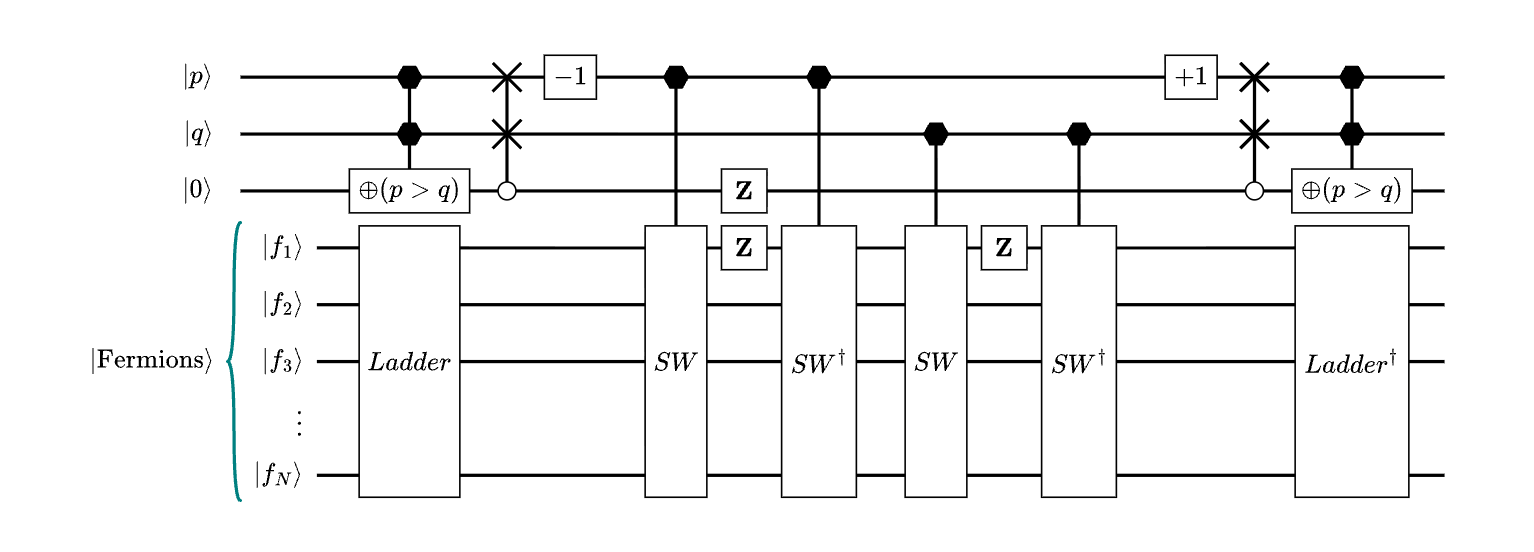},
\end{equation}
and $O_f^{++}$ is
\begin{equation}\label{circ:ofpp}
    \includegraphics[height=4.5cm,valign=c]{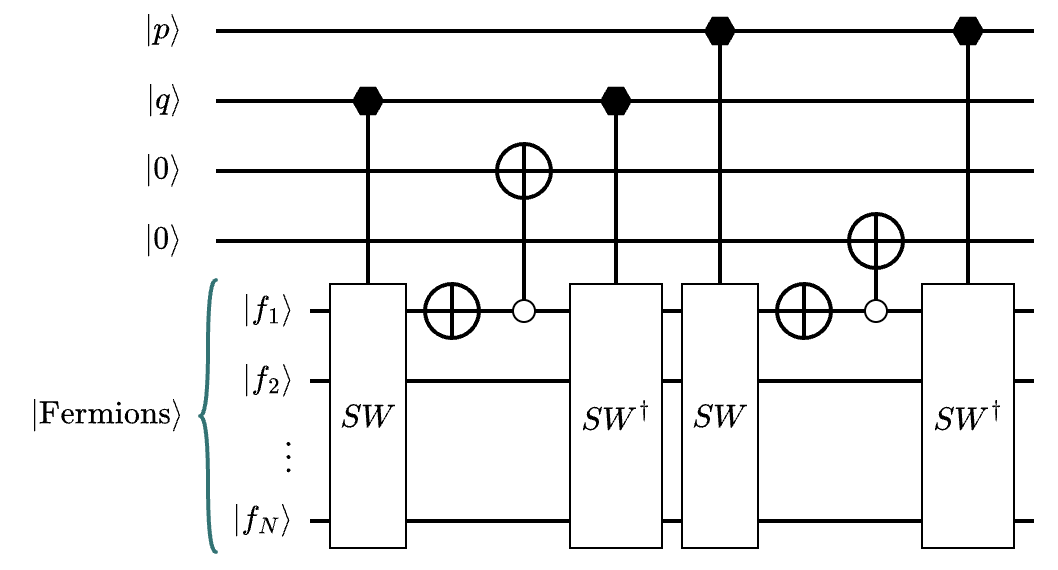}.
\end{equation} 
Unlike $O^{+-}_Z$, $O^{--}_Z$ and $O^{++}_Z$ have an extra $Z$ gate applied to $\ket{(p<q)}$ and $\ket{(p>q)}$, respectively, to impart the negative sign from~\eqref{eq:JW_pp} and~\eqref{eq:JW_mm}. Note that while $O^{++}_f O^{++}_Z$ and $O^{--}_f O^{--}_Z$ share the index ancilla registers, i.e., $\ket{p}\ket{q}$, they have separate flag ancilla qubits. The circuits from~\cite{liu2025blockencodinglowgate} that are analogous to $O^{++}_Z$ / $O^{--}_Z$ and $O^{++}_f$ / $O^{--}_f$ consist of 6 and 8 SwapUps; so our two-body $\sel$ circuit achieves a $\sim 40 \%$ Toffoli count reduction. Finally, by inserting $O^{+-}_f O^{+-}_Z$ between $O^{++}_f O^{++}_Z$ and $O^{--}_f O^{--}_Z$, we obtain the three-body $\sel$ circuit.

We have shown the one, two, and three-body $\sel$ circuits individually. Next, we will describe $\sel$ and $\prep$ circuits that block-encode them together. Let us start with the simpler, phenomenological Hamiltonians that consist of the diagonal one-body term from~\eqref{eq:diag1bd} and the two-body term from~\eqref{eq:NuHamil}. The $\sel$ circuit is given by an open-controlled one-body $\sel$ followed by a closed-controlled two-body $\sel$, i.e.,
\begin{equation}
    (\ketbra{0}{0}\otimes \textsc{Sel}_1 + \ketbra{1}{1}\otimes I)(\ketbra{0}{0}\otimes I + \ketbra{1}{1}\otimes \textsc{Sel}_2) = \ketbra{0}{0}\otimes \textsc{Sel}_1 + \ketbra{1}{1}\otimes \textsc{Sel}_2,
\end{equation}
where $\sel_1$ and $\sel_2$ denote the one and two-body $\sel$ respectively. To control $O_Z$ and $O_f$ in the two-body $\sel$, one only needs to control the operations that are not uncomputed by their inverses, i.e., the $Z$ and (controlled) NOT gates sandwiched by the SwapUps. The $\prep$ circuit prepares the state:
\begin{align}\label{eq:prep2B}
    \prep \ket{0} \mapsto \sum_{p} \sqrt{\frac{t_p}{\lambda}}\ket{0}\ket{0}\ket{p000}\ket{\text{temp}_{l(0,p)}} + \sum_{(pq)\leq (rs)}\sqrt{\frac{V_{pqrs}}{\lambda}}\omega_{pqrs}\ket{1}\ket{+}\ket{pqrs}\ket{\text{temp}_{l(1,pqrs)}},
\end{align}
where the 1-norm $\lambda = \sum_{p}\frac{|t_p|}{2} + \sum_{(pq)\leq (rs)} |V_{pqrs}|$ and
\begin{equation}
    \omega_{pqrs} =
    \begin{cases}
        \sqrt{2} \text{ if }(pq)<(rs), \\
        1 \text{ if }(pq)=(rs), \\
        0 \text{ if }(pq)>(rs),
    \end{cases}
\end{equation}
by applying Alias Sampling~\cite{PhysRevX.8.041015}:
\begin{equation}\label{circ:alias}
    \includegraphics[height=5cm,valign=c]{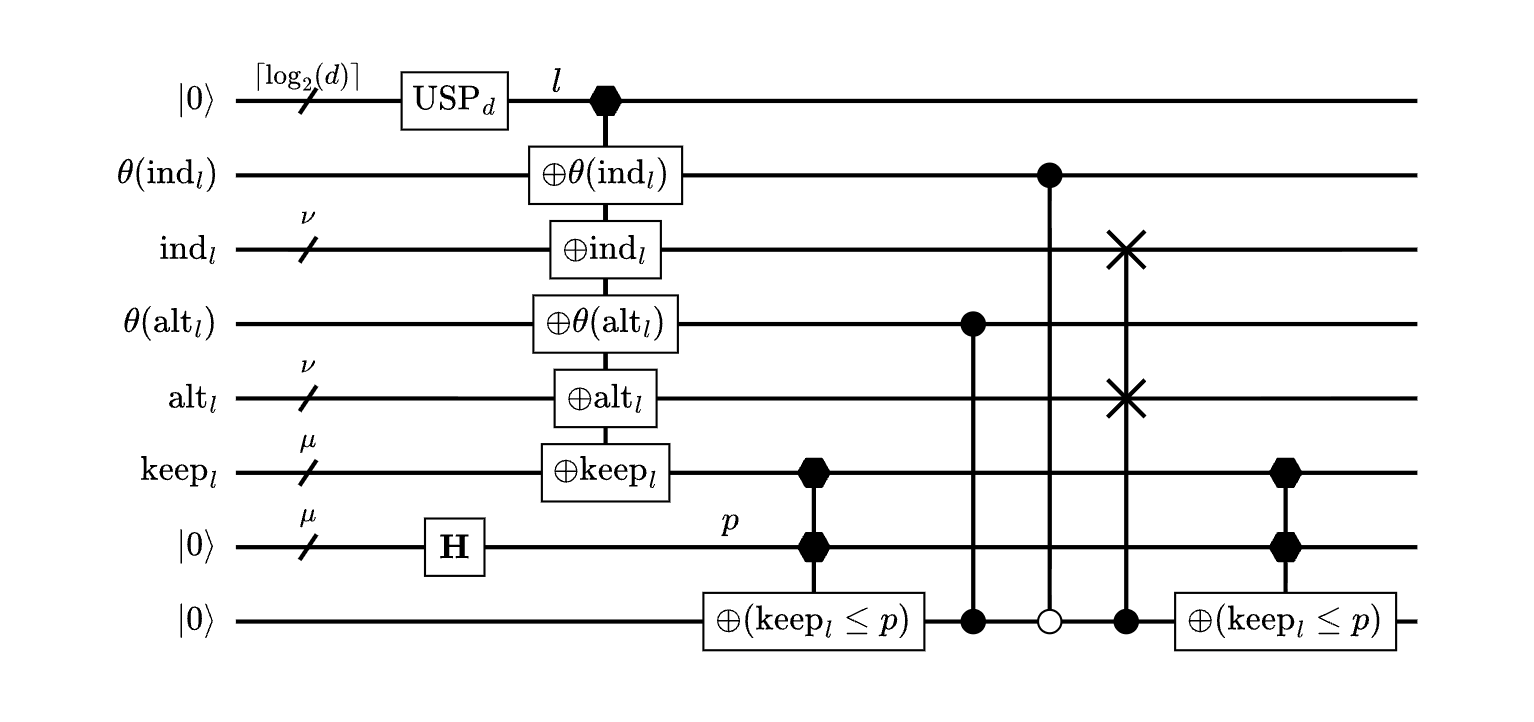},
\end{equation}
followed by a Hadamard gate acting on the second register controlled by the first register. Note that we have largely borrowed the sparse $\prep$ circuit from~\cite{Berry2019qubitizationof} with minor modification. In the Alias Sampling circuit, $\text{USP}_d$ prepares a uniform superposition~\footnote{We do this using the amplitude amplification circuit in figure 12 in~\cite{PhysRevX.8.041015} or figure 3 in~\cite{PRXQuantum.2.030305}.}: $\frac{1}{\sqrt{d}}\sum_{l=1}^d\ket{l}$ where $d$ is the number of nonzero $t_p$ plus that of nonzero $V_{pqrs}$ with $(pq) \leq (rs)$; $(pq)$ is a composite index obtained by vectorizing the indices $p$ and $q$. The subsequent operation is a classical table look-up implemented using a QROM: $\ket{\text{ind}_l}$ stores the label that distinguishes between the one and two terms, i.e., first qubit in~\eqref{eq:prep2B}, and the tensor indices, i.e., third register in~\eqref{eq:prep2B}. As such, $\ket{\text{ind}_l}$ consists of $\nu = 4\lceil \log_2(N)\rceil + 1$ qubits. Here $l$ can be viewed as the bijective function between the integers $[1..d]$ and the labels. $\ket{\theta(\text{ind}_l)}$ stores the sign of the one or two-body element, as indicated by the labels; e.g., if the $l$th index is for a two-body tensor, then $\ket{\theta(\text{ind}_l)}\ket{\text{ind}_l} = \ket{\text{sign}(V_{pqrs})}\ket{1}\ket{pqrs}$. $\ket{\theta(\text{alt}_l)}\ket{\text{alt}_l}$ is of the same size, and $\ket{\text{keep}_l}$ is a $\mu$-bit register so that the amplitudes of the final prepared state are $\mu$-bit approximations of the target amplitudes. Both $\ket{\text{alt}_l}$ and $\ket{\text{keep}_l}$ are classically computed. The remaining operations swap $\ket{\text{ind}_l}$ and $\ket{\text{alt}_l}$ with probability $\text{keep}_l / 2^\mu$, and the corresponding negative signs are applied depending on whether a swap was applied. Note that the QROM that implements the classical data look-up is the most expensive routine in $\prep$ and $\sel$. To this end, we employ the clean-ancilla-assisted QROM~\cite{Berry2019qubitizationof,Low2024tradingtgatesdirty}, which enables Toffoli-reduction at the cost of clean ancilla qubits; the Toffoli and qubit costs of the computation and uncomputation of this QROM are stated in theorems 2 and 3 of~\cite{Berry2019qubitizationof}, respectively. How $\mu$ is determined in relation to the algorithmic error is discussed in appendix~\ref{app:query}. Finally, the $\ket{\text{temp}_l}$ register stores a garbage state that is harmless and does not introduce any errors, as it is not acted on by $\sel$ and is uncomputed by $\prep^\dagger$. To understand the Alias Sampling method better, please read section III.D in~\cite{PhysRevX.8.041015}.

Now we consider the full Hamiltonian with a three-body term, e.g., a chiral EFT Hamiltonian, from~\eqref{eq:NuHamil}. The $\prep$ circuit prepares
\begin{align}\label{eq:prep3B}
    &\prep \ket{0} \mapsto \nonumber \\
    &\sum_{p\leq q} \text{sign}(t_{pq})\sqrt{\frac{|t_{pq}|}{\lambda}}\theta_{pq} \ket{00}\ket{+}\ket{00pp00}  + \sum_{(pq)\leq (rs)}\text{sign}(V_{pqrs})\sqrt{\frac{|V_{pqrs|}}{\lambda}}\omega_{pqrs} \ket{01}\ket{+}\ket{pq00rs} \nonumber \\
    &+ \sum_{(pqr)\leq (stu)}\text{sign}(W_{pqrstu})\sqrt{\frac{|W_{pqrstu|}}{\lambda}} \gamma_{pqrstu}\ket{10}\ket{+}\ket{pqrstu},
\end{align}
where
\begin{equation}
    \theta_{pq} =
    \begin{cases}
        \sqrt{2} \text{ if }p<q, \\
        1 \text{ if }p=q, \\
        0 \text{ if }p>q,
    \end{cases}
\end{equation}
and 
\begin{equation}
    \gamma_{pqrstu} =
    \begin{cases}
        \sqrt{2} \text{ if }(pqr)<(stu), \\
        1 \text{ if }(pq)=(rs), \\
        0 \text{ if }(pq)>(rs),
    \end{cases}
\end{equation}
by applying Alias Sampling, followed by a Hadamard on $\ket{0}$ (third register). Note that we have left out the $\text{temp}_l$ registers for brevity. Furthermore, $\ket{\text{ind}_l}$ stores the first and third register and thus, in~\eqref{circ:alias}, $\nu = 6 \lceil \log_2(N)\rceil + 2$. $d$ is the number of nonzero upper triangular and diagonal elements in the one, two, and three-body term, and the 1-norm $\lambda = \sum_{pq} |t_{pq}| + \sum_{pqrs} |V_{pqrs}| +\sum_{pqrstu} |W_{pqrstu}|$. The $\sel$ circuit is given by
\begin{equation}
    \includegraphics[height=4.5cm,valign=c]{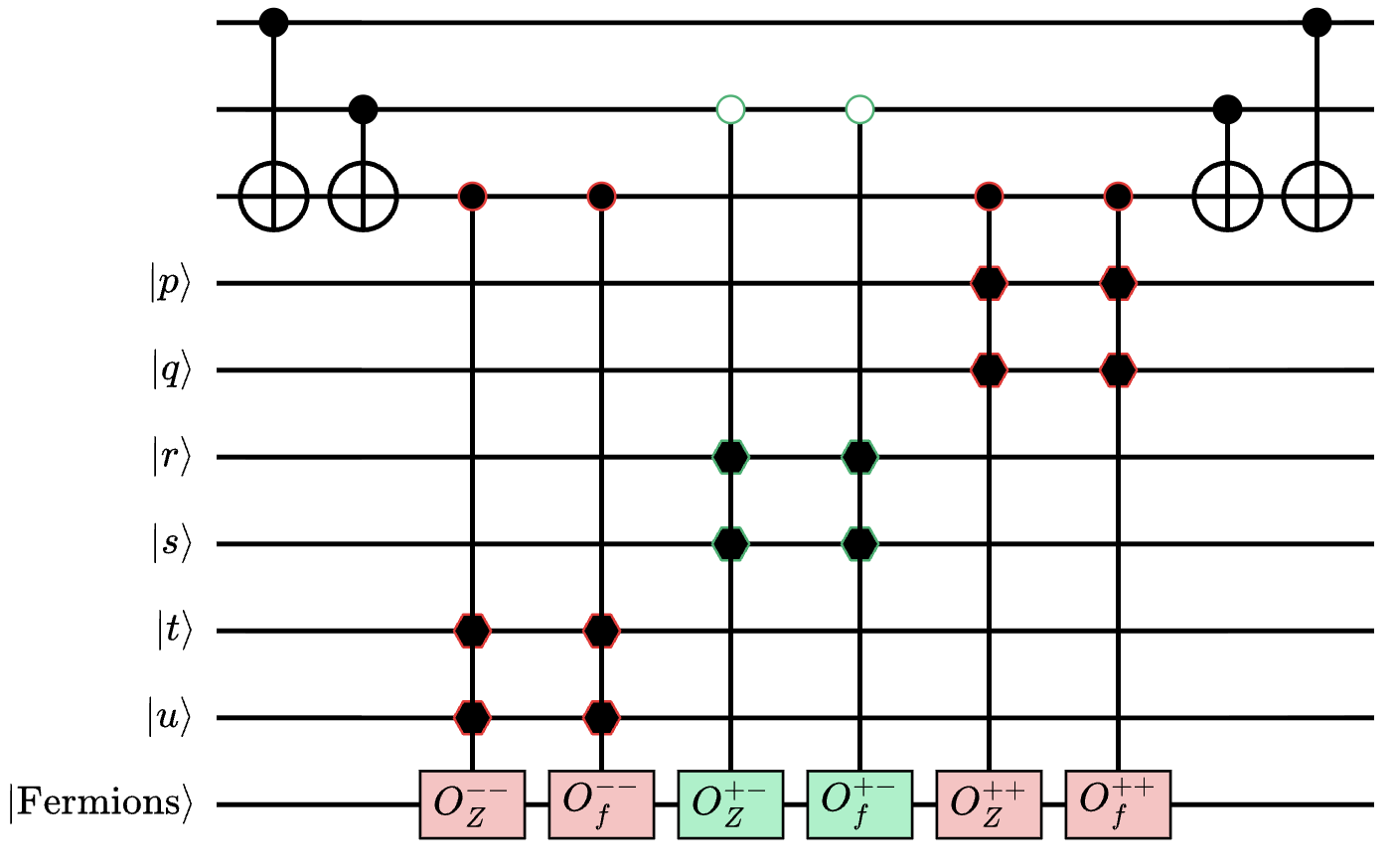},
\end{equation}
where the first two qubits are the first two qubits in the state prepared by $\textsc{Prep}$, the CNOTs compute the XOR of the first two qubits into the third qubit (a new ancilla not involved in $\prep$) before uncomputing it, and the flag ancilla qubits in $O_Z$ and $O_f$ are hidden. As such, the red operations are applied when the first two qubits are either $\ket{01}$ or $\ket{10}$, and the green operations are applied when the first two qubits are either $\ket{00}$ or $\ket{01}$. $\textsc{Prep}^\dagger\cdot \textsc{Sel} \cdot \textsc{Prep}$ block-encodes the Hamiltonian because the green operations perform $\textsc{Sel}_1$, the red operations perform $\textsc{Sel}_2$, and the green and red operations combined perform $\textsc{Sel}_3$. Note that now the block-encoded one-body term is $\sum_p \frac{1-Z_p}{2}$ instead of $\sum_p Z_p$. We want to highlight that this circuit has roughly the same size and cost as the $\sel$ for just the three-body term, and its costs are roughly half those of the $\sel$ of one, two and three-body term combined as a LCU, which is the approach proposed in~\cite{liu2025blockencodinglowgate}. Combined with the $\sim 40\%$ Toffoli reduction in $O^{\pm\pm}_Z$ and $O^{\pm\pm}_f$ over the analogous circuits in~\cite{liu2025blockencodinglowgate} discussed above, our $\sel$ implementation achieves a $\sim 70\%$ Toffoli-reduction over that in~\cite{liu2025blockencodinglowgate}.

\subsection{Quantum resource estimates}
\label{sec:qre}

In this section, we apply our circuits to compile the qubitization-based phase estimation algorithms for the $^{24}$Mg~\cite{PhysRevC.74.034315}, $^{32}$Mg~\cite{PhysRevC.79.014310}, and $^{219}$At~\cite{PhysRevC.43.602} shell-model Hamiltonians, as well as a Chiral EFT, no-core shell-model Hamiltonian suitable for light nuclei $\sim$40Ca, and report the resulting quantum resource estimates in terms of Toffoli and logical qubit count. For our algorithms, we set the target simulation error to be $\epsilon=0.01$ in the same unit of the Hamiltonian matrix element, i.e., MeV. When it comes to simulation accuracies in nuclear physics, there is no golden standard that is analogous to ``chemical accuracy" in chemistry calculations. In part, this is because unlike in chemistry where there is only one choice of interaction, i.e., Coulomb interaction, even for a given nucleus, there could be various nuclear interactions to choose from and include in its Hamiltonian, and the ground-state energies can differ more than 0.01 between different Hamiltonians~\cite{PhysRevC.74.034315,Navratil_2009}. In this sense, $\epsilon=0.01$ is a reasonable choice, but we encourage further studies to rigorously derive $\epsilon$-values that are sufficient for nuclear simulations.

First, we address the phenomenological Hamiltonians. We consider two strategies: (i) We apply quantum phase estimation directly to entire Hamiltonians. (ii) We exploit the fact that the Hamiltonians are $M$-scheme Hamiltonians, which are expressed in a specific basis where the two-body term are block-diagonal with each block labeled by the total angular momentum projection $M$~\cite{Whitehead1977}, which loosely speaking, originate from adding together pairs of single-particle quantum numbers. In particular, instead of simulating the whole Hamiltonian, we apply quantum phase estimation separately to every $M$-block of the two-body term, combined with the one-body term~\footnote{In principle, we can simulate a part, instead of the whole, of the one-body term, which includes only the single-particle basis states that are relevant to the simulated two-body $M$-block. However, we expect the cost reduction from doing this will be small because the overall costs are dominated by the two-body term.}. If one is interested in using such an algorithm in the context of ground-state energy estimation, one can pick out the lowest quantum phase estimation output classically since the number of $M$-blocks is very manageable. For example, $^{24}$Mg has 11, $^{32}$Mg has 15, and $^{219}$At has 29 $M$-blocks. The advantage of approach (ii) is that the qubit count will be determined by the size of the largest $M$-block, which is always smaller than the entire Hamiltonian, i.e., sum of all $M$-blocks, and thus, it is expected for approach (ii) to incur a lower qubit count. Furthermore, the physical size of a fault-tolerant quantum computer, i.e., code distance of the error correction code and number of physical qubits, which can execute a given algorithm is determined by the number of qubits and gates in the largest circuit executed coherently as part of the algorithm. In other words, the size requirement of a fault-tolerant quantum computer for approach (ii) will be determined by the largest $M$-block, instead of the entire Hamiltonian.

In particular, for approach (i), we explore the space-time trade-off between Toffolis and qubits by tweaking the parameter $k$ (see theorems 2 and 3 of~\cite{Berry2019qubitizationof}) of the ancilla-assisted QROM, and report the resulting resource estimates in the first three lines of tables~\ref{tb:24Mg},~\ref{tb:32Mg}, and~\ref{tb:219At}, whereas the resource estimates for approach (ii) are in the last boldfaced line of the tables. Note that the largest coherent QPE circuit for $^{24}$Mg, $^{32}$Mg, and $^{219}$At consist of 2.08e8, 1.95e10, and 1.44e10, respectively, with qubit counts stated in the bottom right cells of tables~\ref{tb:24Mg},~\ref{tb:32Mg}, and~\ref{tb:219At}. From the resource estimates, it is clear that approach (ii) is more resource-efficient, in terms of Toffoli and logical qubit counts, code distance and thus physical qubit count, than approach (i). Furthermore, compared to the best algorithms for simulating the Femoco Hamiltonian, which has slightly more orbitals than the $^{32}$Mg and much fewer orbitals than $^{219}$At~\footnote{Femoco has 76 orbitals, $^{32}$Mg has 64 orbitals, and $^{219}$At has 102 orbitals.}, the largest coherent QPE circuit for $^{32}$Mg requires half the number of qubits and about an order of magnitude more Toffoli gates, and $^{219}$At requires roughly the same number of qubits and again about an order of magnitude more Toffoli gates~\cite{pb2g-j9cw,yngp-5fpm}. We provide the Hamiltonian data, i.e., 1-norms and number of non-zero elements, in appendix~\ref{app:nuHam}.

\begin{table}[!ht]
\centering
\begin{subtable}{0.32\textwidth}
\centering
\begin{tabular}{|ll|}
\hline
\multicolumn{2}{|c|}{$^{24}$Mg}                           \\ \hline
\multicolumn{1}{|l|}{$\#$ Toffolis}   & $\#$ Qubits  \\ \hline
\multicolumn{1}{|l|}{1.07e9}         & 608         \\ \hline
\multicolumn{1}{|l|}{1.09e9}         & 349         \\ \hline
\multicolumn{1}{|l|}{1.36e9}         & 220          \\ \hline
\multicolumn{1}{|l|}{\textbf{9.52e8}} & \textbf{208} \\ \hline
\end{tabular}
\caption{}
\label{tb:24Mg}
\end{subtable}
\hfill
\begin{subtable}{0.32\textwidth}
\centering
\begin{tabular}{|ll|}
\hline
\multicolumn{2}{|c|}{$^{32}$Mg}                            \\ \hline
\multicolumn{1}{|l|}{$\#$ Toffolis}    & $\#$ Qubits  \\ \hline
\multicolumn{1}{|l|}{1.16e11}          & 2640         \\ \hline
\multicolumn{1}{|l|}{1.20e11}          & 1393         \\ \hline
\multicolumn{1}{|l|}{3.44e11}          & 770          \\ \hline
\multicolumn{1}{|l|}{\textbf{9.38e10}} & \textbf{746} \\ \hline
\end{tabular}
\caption{}
\label{tb:32Mg}
\end{subtable}
\hfill
\begin{subtable}{0.32\textwidth}
\centering
\begin{tabular}{|ll|}
\hline
\multicolumn{2}{|c|}{$^{219}$At}                            \\ \hline
\multicolumn{1}{|l|}{$\#$ Toffolis}    & $\#$ Qubits   \\ \hline
\multicolumn{1}{|l|}{3.07e11}          & 8.42e4        \\ \hline
\multicolumn{1}{|l|}{4.80e11}          & 2.12e4        \\ \hline
\multicolumn{1}{|l|}{1.71e12}          & 5442          \\ \hline
\multicolumn{1}{|l|}{\textbf{2.90e10}} & \textbf{1486} \\ \hline
\end{tabular}
\caption{}
\label{tb:219At}
\end{subtable}
\caption{\textbf{Quantum resource estimates for $M$-scheme shell-model Hamiltonians of (a) $^{24}$Mg, (b) $^{32}$Mg, and (c) $^{219}$At.} The numbers in bold are derived by applying phase estimation to one sub-block, labeled by the $M$ quantum number, of the Hamiltonians at a time, whereas the rest are for simulations of entire Hamiltonians without any partitioning.}
\label{tb:2body}
\end{table}

Let us proceed to the no-core shell-model Hamiltonians with up to three-body interactions derived from Chiral EFT. In principle, the same symmetry in the total angular momentum projection $M$ used to reduce the cost of block encoding the phenomenological Hamiltonians applies to these models, and an algorithm for doing so is outlined in appendix~\ref{app:chiral_block_decomp}. However, this strategy does not provide the same improvements in resource costs that were observed for the phenomenological Hamiltonians.

The reason for this is that the Hamiltonian is partitioned into sectors with distinct quantum numbers $M_3$ that block-diagonalize the three-body terms. However, the dominant terms in the Hamiltonian both in terms of the number of non-zero matrix elements and $1-$norm are the two-body interactions. Since many of the two-body terms need to be included in the simulation of any given $M_3$ sector, the result is that partitioning yields only a small decrease in qubit count whilst incurring a significant increase in the number of Toffolis required.

\begin{table}[ht!]
\begin{tabular}{|c|c|}
\hline
$\#$ Toffolis & $\#$ Qubits \\
\hline
4.11e14 & 6.78e4 \\
\hline
9.46e14 & 1.71e4 \\
\hline
3.49e15 & 4487 \\
\hline
6.99e15 & 2376 \\
\hline
\end{tabular}
\caption{\textbf{Quantum resource estimates for a no-core shell-model derived from chiral EFT with $e_{3\max} =10$.} Rows of resources correspond to different configurations of space-time tradeoff parameters for the state preparation subroutine.}
\end{table}
\label{tab:3body_qres}

The quantum resources required to simulate the no-core shell-model Hamiltonian with $e_{3\max}=8$ are given in table~\ref{tab:3body_qres}. We provide the Hamiltonian data, i.e., 1-norms and number of non-zero elements, in appendix~\ref{app:nuHam}.

The estimates are substantially higher than for the phenonenological models, reflecting the much larger size of the Hamiltonian analyzed. Notably, the most problematic terms are the two-body interactions, rather than the three-body. The number of non-zero upper-diagonal elements for the two-body interactions in the no-core Hamiltonian is approximately $5\times10^{7}$, compared with $7\times10^4$ for the two-body terms in the largest phenomenological model and $2.7\times10^5$ for the three-body terms. Similarly, the one-norm for the two-body interactions is approximately $5.9\times10^6$, compared with $4.2\times10^4$ for the two-body interactions in the phenomenological model of $^{32}$Mg and 2099.76 for the three-body terms.

While the quantum resources required for the simulation of no-core shell-model Hamiltonians appear daunting, it should be noted that they come from an entirely unfactorized Hamiltonian, block encoded using techniques devised for electronic structure problems. In that setting, tensor factorization techniques such as double factorizaion, block-invariant symmetry shift and tensor hypercontraction have improved qubit and gate costs by orders of magnitude. It is not unreasonable to expect that similar gains may be available in the nuclear setting.

% For the Hamiltonian with $e_{3\max} =8$, \ins{QREs in table + discussion around them; QREs for simulating the whole Hamiltonian vs block-based approach}. For the Hamiltonian with $e_{3\max} =10$, \ins{QREs in table + discussion around them; QREs for simulating the whole Hamiltonian vs block-based approach}. We provide the Hamiltonian data, i.e., 1-norms and number of non-zero elements, in appendix~\ref{app:nuHam}.

\section{Discussion}
\label{sec:discussion}

In this work, we have constructed fault-tolerant energy estimation algorithms and provided the first, to our knowledge, quantum resource estimates for simulating atomic nuclei. In particular, we considered $^{24}$Mg~\cite{PhysRevC.74.034315}, $^{32}$Mg~\cite{PhysRevC.79.014310}, and $^{219}$At~\cite{PhysRevC.43.602} shell-model Hamiltonians, and no-core shell-model Hamiltonians, including three-body forces derived from chiral EFT, which are applicable to lighter nuclei up to $^{40}$Ca or so~\cite{Miyagi_2023}. Our algorithms are based on qubitization, which underlie many quantum algorithms for simulating fermionic systems in chemistry and condensed matter~\cite{PhysRevX.8.041015,Berry2019qubitizationof,PhysRevResearch.3.033055,Kan_2025,PRXQuantum.2.030305,pb2g-j9cw,yngp-5fpm}. Notably, we find the resource estimates for simulating the classical difficult-to-simulate $^{32}$Mg and $^{219}$At to be close to the best estimates for simulating the Femoco molecule~\cite{yngp-5fpm,pb2g-j9cw}.

For our algorithm, we construct the input model, namely, the block-encoding of the Hamiltonians, by adapting a state-preparation protocol, i.e., $\prep$, that exploits the Hamiltonians' sparsity designed for chemistry Hamiltonians~\cite{Berry2019qubitizationof} and combining it with encoding circuits of individual monomials of fermionic operator, i.e., $\sel$, from~\cite{liu2025blockencodinglowgate} that we optimize; when applied to Hamiltonians with and without the three-body term, our optimizations lead to a $70\%$ and $40\%$ reduction, respectively, in Toffoli count over the comparable implementations in~\cite{liu2025blockencodinglowgate}. Further reported in~\cite{liu2025blockencodinglowgate} is an alternative block-encoding method that combines $\sel$ with a different state-preparation protocol called the amplitude oracle, leading to algorithms for simulating two-body Hamiltonians that scale like $O(N^4 \lambda_{\max})$ and $O(\eta^2 N^2 \lambda_{\max})$ where $\eta$, $N$, and $\lambda_{\max}$ are the particle number, orbital number, and max norm of the Hamiltonian, respectively. Our algorithm scales like $O(d \lambda_\text{one})$ where $d$ is roughly the number of monomials of fermionic operators, i.e., sparsity, and $\lambda_\text{one}$ is the LCU 1-norm. For the shell-model Hamiltonians considered in this work, $\eta \sim 0.1 N$, $d \sim 0.01 N^4$ and $N^4 \lambda_{\max} \gg \lambda_\text{one}$, which means our algorithm scales better.

By borrowing certain techniques from chemistry algorithms, we are able to achieve resource estimates that are close to recent estimates obtained for simulating electronic structure~\cite{PhysRevX.8.041015,Berry2019qubitizationof,PhysRevResearch.3.033055,PRXQuantum.2.030305,pb2g-j9cw,yngp-5fpm}. However, we certainly have not exhausted the lessons from chemistry that could potentially be applied to nuclear-physics settings in future work. For example, if one chooses to apply our algorithm to estimate ground-state energies, one will need to construct an initial state with a good overlap with the ground state. This is generally not an easy problem, and in the worst case, is harder than the energy estimation algorithm. This is largely resolved in chemistry, where state-of-the-art methods have comparable costs to the energy estimation algorithm~\cite{PRXQuantum.6.020327,PRXQuantum.5.040339}. Briefly, these methods first obtain a MPS ansatz via classical simulation and then convert it into an optimized quantum circuit. Unfortunately, MPS (or equivalently, DMRG) classical simulations are much less developed for nuclear structure calculations, with a limited number of publications on this topic~\cite{PhysRevLett.97.110603,PhysRevC.79.014304,TICHAI2023138139,TICHAI2024138841,Dukelsky_2004,PhysRevC.78.041303,PhysRevC.92.051303,PhysRevC.65.054319}, especially when compared to chemistry (see~\cite{10.1063/1.4939000,baiardi2019large} and references therein). Moreover, the MPS nuclear simulations that we are aware of do not simulate the three-body term. Therefore, developing MPS classical simulation will likely be a bottleneck for the initial-state problem, if one adopts the MPS approach. Alternatively, one could consider dissipative state-preparation methods, which have shown early signs of success in chemistry and condensed matter~\cite{watts2026rapiddissipativegroundstate,PhysRevResearch.6.033147,zhan2025rapidquantumgroundstate,_m_d_2025}, though remain untested in nuclear physics. As in MPS-based methods, theoretical and empirical evidences are needed to understand the performance and costs of such methods in nuclear settings.

Another chemistry-inspired direction that is worth exploring is tensor factorization techniques, which has been the main driving force behind various recent optimizations of chemistry algorithms~\cite{Berry2019qubitizationof,Motta_2021, PhysRevResearch.3.033055,PRXQuantum.2.030305,pb2g-j9cw,yngp-5fpm}. However, such techniques rely on properties of electronic-structure Hamiltonians that are absent in nuclear Hamiltonians. For example, a so-called eightfold symmetry, which essentially is an chemistry-specific orbital-permutation symmetry, in electronic-structure Hamiltonians~\cite{Motta_2021, Berry2019qubitizationof} is exploited in such techniques. This symmetry manifests in the symmetry between factor matrices that are outputted from factorization algorithms, and in turn, is reflected in the symmetry of the multiplexed rotations in the quantum circuit implementation of factorization-based algorithms. Furthermore, an important factor in the performance of such techniques is that the two-body term in electronic-structure Hamiltonians, upon application of a singular value decomposition, retains only $O(N)$ number of non-trivial singular values; singular value decomposition is often the first step of many tensor factorization routines in chemistry and the $O(N)$ scaling plays a crucial role in efficiency of factorization-based algorithms. Unfortunately, nuclear Hamiltonians can retain $O(N^2)$ singular values or even be full-rank, which might be due to the long-range nature of nuclear interactions~\cite{PhysRevResearch.6.043331,PhysRevC.99.034320}. Therefore, it is unlikely tensor factorization techniques with chemistry origins can be applied, without appropriate modifications, to nuclear Hamiltonians. A lower-hanging fruit than tensor factorization is the addition of symmetry shifts to Hamiltonians, which can often reduce their 1-norm and thus gate counts of chemistry simulations~\cite{doi:10.1021/acs.jctc.4c00352,yngp-5fpm,PhysRevA.110.022420}.

Finally, a promising future direction is to apply our circuits to construct time-evolution algorithms for simulating nuclear reaction problems that contain three-body interactions~\cite{PhysRevC.87.034326,PhysRevC.79.044606,navratil2016unified}.

\section*{Acknowledgements}
The quantum circuit diagrams in this work are generated using Circuit Designer (freely available at \url{http://circuits.psiquantum.com}) from PsiQuantum's Construct software suite.  Support from the UK STFC under grant ST/Y000358/1 is acknowledged. 

UK Ministry of Defence \copyright Crown Owned Copyright 2026/AWE.

\section*{Author Contributions}
The author list is ordered alphabetically. J. B., N. G., A. K., and P. S. contributed to the conception of the work. A. K. designed the quantum algorithms and circuits. M. G., S. G., and A. K. performed the resource estimation. L. L. R., C. S., and P. S. generated the nuclear Hamiltonians. All authors contributed to the writing and editing of the manuscript.

\bibliography{ref}
\bibliographystyle{apsrev4-2}

\appendix

\section{Error budgeting and query count of walk operator}
\label{app:query}

As is commonly done~\cite{PhysRevX.8.041015,Berry2019qubitizationof,Kivlichan2020improvedfault,Campbell_2022,Kan_2025,PRXQuantum.2.030305}, we take the quantum phase estimation error to be the root-mean-square error of the estimated phase $\Delta \phi$, which is related to the energy error budget $\Delta E$ via
\begin{equation}
    \Delta E = \lambda \Delta \cos{(\phi)} \leq \lambda \Delta \phi.
\end{equation}
Recall that the Hamiltonian $H = \sum_{i=1}^L c_i O_i$ and $\lambda = \sum_{i=1}^L |c_i|$.

First, we divide the error budget into two sources: the Holevo variance~\cite{PhysRevX.8.041015} of the phase estimation outcome $\epsilon_\text{QPE} = \frac{\pi}{2^{m+1}}$ and the error from constructing the walk operator $\epsilon_\mathcal{W}$, such that
\begin{equation}
    \left( \frac{\Delta E}{\lambda}\right)^2 \leq (\Delta \phi)^2 = \epsilon_\text{QPE}^2 + \epsilon^2_\mathcal{W},
\end{equation}
and
\begin{equation}
    \epsilon_\text{QPE} = \sqrt{x}\frac{\Delta E}{\lambda}, \: \epsilon_\mathcal{W} = \sqrt{1-x}\frac{\Delta E}{\lambda},
\end{equation}
where $0<x<1$. Then, we can compute the sufficient number of phase qubits, which stores the estimated phase, as
\begin{equation}
    \frac{\pi}{2^{m+1}} = \frac{\sqrt{x}\Delta E}{\lambda} \implies m = \left \lceil \log_2{\Bigg(\frac{\pi \lambda}{2\sqrt{x}\Delta E}\Bigg)} \right \rceil.
\end{equation}
The number of queries to the walk operator in the quantum phase estimation algorithm is then $Q_\mathcal{W}=2^m$. 

$\epsilon_\mathcal{W}$ is further divided into alias sampling error $\epsilon_\text{AS}$ and rotation synthesis error $\epsilon_\text{r}$ such that 
\begin{equation}
    \epsilon_\text{AS} = y \epsilon_\mathcal{W}, \: \epsilon_\text{r} = (1-y) \epsilon_\mathcal{W},
\end{equation}
where $0<y<1$. Using $\epsilon_\text{AS}$, we can determine the size of the registers, i.e., $\mu$, that are used to compute the swapping probability in the alias sampling circuit. In particular, we start with the relation $\frac{1}{2^\mu} \leq \frac{\delta L}{\lambda}$, taken almost verbatim from equation 35 in~\cite{PhysRevX.8.041015}, where the right hand side is given by equation A9 in~\cite{PhysRevX.8.041015}:
\begin{equation}
    \frac{\delta L}{\lambda} \geq \frac{\epsilon_\text{AS}}{1+\epsilon_\text{AS}^2} \left (1-\frac{\|H \|^2}{\lambda^2} \right ).
\end{equation}
Then, one can set
\begin{align}
    \mu &= \log_2{\left ( \epsilon_\text{AS}^{-1} \right )} + \log_2{\left ( 1+\epsilon_\text{AS}^2\right )} - \log_2{\left ( 1-\frac{\|H \|^2}{\lambda^2}\right )} \label{eq:mu_val1} \\
    &\approx \lceil \log_2{\left ( \epsilon_\text{AS}^{-1} \right )} \rceil \label{eq:mu_val2},
\end{align}
where the last two terms in~\eqref{eq:mu_val1} are constants that are less than 1 and hence the second line~\cite{PhysRevX.8.041015}.

\section{Details of nuclear Hamiltonians}
\label{app:nuHam}

\subsection{\texorpdfstring{$^{24}$Mg}{}}
The $^{24}$Mg case makes use of the USDB Hamiltonian~\cite{PhysRevC.74.034315}. This acts in the $sd$ major shell, defined as those levels between the magic numbers 8 and 20. Thus the $^{16}$O nucleus is considered as an inert core with valence protons and neutrons each able to occupy the set of orbitals $\{0d_{5/2},0s_{1/2},0d_{3/2}\}$. The two body matrix elements and single particle energies were emprically fitted to 608 data points across 77 nuclei in the $sd$ shell. 

The empirical fit yields a baseline set of matrix elements, which are augmented by a mass dependence, auch that the tabulated values are suitable for mass number $A=18$ nuclei, with two-body matrix elements related to the $A=18$ values by

\begin{equation}
    V_{JT}(a,b,c,d,A) = \left(\frac{18}{A}\right)^p V_{JT}(a,b,c,d,18),
\end{equation}

with $p=3$.  Here, the $a,b,c,d$ indices refer to the spherical shell model basis in the $JT$-coupled basis -- i.e. the set $\{0d_{5/2},0s_{1/2},0d_{3/2}\}$.

\begin{table}[!ht]
\begin{tabular}{|c|c|c|}
\hline
$M$     & $\sum_{pqrs}|V_{pqrs}|$ & $|\{V_{pqrs} : |V_{pqrs}| > 0 \: \& \: (pq) \leq (rs)\}|$ \\ \hline
0                            & 462.37                 & 592                                                 \\ \hline
$\pm 1$                     & 339.17                 & 453                                                 \\ \hline
$\pm 2$                       & 211.51                 & 261                                                 \\ \hline
$\pm 3$                       & 90.72                  & 73                                                  \\ \hline
$\pm 4$                      & 29.23                  & 16                                                  \\ \hline
$\pm 5$                       & 3.96                   & 1                                                   \\ \hline
\end{tabular}
\caption{1-norm and number of non-zero upper-diagonal elements of two-body terms per $M$-block of the Hamiltonian. The 1-norm and number of non-zero upper-diagonal elements of the one-body term are 76.83 and 24, respectively.}
\label{tb:24Mg_info}
\end{table}

\subsection{\texorpdfstring{$^{32}$Mg}{}}
For $^{32}$Mg, the \textit{SDPF-U} interaction~\cite{PhysRevC.79.014310} is used.  This operates in a valence space consisting of the $sd$ shell for protons, and the combined $sdpf$ shell for neutrons. This interaction is a development of the previous SPDF-NR interaction~\cite{retamosa_shell_1997,nummela_spectroscopy_2001}, derived from a mixture of emprical and theoretical considerations.  It is a widely-used interaction in this model space.

\begin{table}[!ht]
\begin{tabular}{|c|c|c|}
\hline
$M$      & $\sum_{pqrs}|V_{pqrs}|$ & $|\{V_{pqrs} : |V_{pqrs}| > 0 \: \&  \: (pq) \leq (rs)\}|$ \\ \hline
0                             & 14342.75               & 10288                                               \\ \hline
$\pm 1$                       & 12056.90               & 8486                                                \\ \hline
$\pm 2$                       & 8338.97                & 5653                                                \\ \hline
$\pm 3$                       & 4609.31                & 2322                                                \\ \hline
$\pm 4$                       & 2035.54                & 776                                                 \\ \hline
$\pm 5$                       & 692.88                 & 119                                                 \\ \hline
$\pm 6$                       & 129.30                 & 21                                                  \\ \hline
$\pm 7$                       & 0.21                   & 1                                                   \\ \hline
\end{tabular}
\caption{1-norm and number of non-zero upper-diagonal elements of two-body terms per $M$-block of the Hamiltonian. The 1-norm and number of non-zero upper-diagonal elements of the one-body term are 384.56 and 64, respectively.}
\label{tb:32Mg_info}
\end{table}

\subsection{\texorpdfstring{$^{219}$At}{}}
Our calculation for $^{219}$At (85 protons and 134 neutrons) uses the KHPE~\cite{PhysRevC.43.602} interaction.  KHPE is an effective interaction based on the renomalization of a free nucleon-nucleon interaction to act in the space beyond a $^{208}$Pb core. The valence space for this interaction is comprised of six proton orbitals (0$h_{9/2}$, 1$ f_{7/2}$, 1$f_{5/2}$, 2$p_{3/2}$, 2$p_{1/2}$, 0$i_{13/2}$) that can accommodate valence protons between $Z$ = 82 and 126 and seven neutron orbitals (0$i_{11/2}$, 1$g_{9/2}$, 1$g_{7/2}$, 2$d_{5/2}$, 2$d_{3/2}$, 3$s_{1/2}$, 0$j_{15/2}$) that can accommodate valence neutrons between $N$ = 126 and 184.

\begin{table}[!ht]
\begin{tabular}{|c|c|c|}
\hline
$M$       & $\sum_{pqrs}|V_{pqrs}|$ & $|\{V_{pqrs} : |V_{pqrs}| > 0 \: \& \: (pq) \leq (rs)\}|$ \\ \hline
0                             & 6487.70                & 22194                                               \\ \hline
$\pm 1$                       & 693.10                 & 13498                                               \\ \hline
$\pm 2$                       & 573.19                 & 11438                                               \\ \hline
$\pm 3$                       & 435.82                 & 8781                                                \\ \hline
$\pm 4$                       & 321.70                 & 6120                                                \\ \hline
$\pm 5$                       & 226.25                 & 3909                                                \\ \hline
$\pm 6$                       & 152.70                 & 2312                                                \\ \hline
$\pm 7$                       & 100.01                 & 1267                                                \\ \hline
$\pm 8$                       & 60.33                  & 617                                                 \\ \hline
$\pm 9$                       & 35.31                  & 265                                                 \\ \hline
$\pm 10$                      & 19.01                  & 99                                                  \\ \hline
$\pm 11$                      & 9.65                   & 31                                                  \\ \hline
$\pm 12$                      & 4.47                   & 9                                                   \\ \hline
$\pm 13$                      & 1.93                   & 3                                                   \\ \hline
$\pm 14$                      & 0.96                   & 1                                                   \\ \hline
\end{tabular}
\caption{1-norm and number of non-zero upper-diagonal elements of two-body terms per $M$-block of the Hamiltonian. The 1-norm and number of non-zero upper-diagonal elements of the one-body term are 253.49 and 102, respectively.}
\label{tb:219At_info}
\end{table}

\subsection{Chiral EFT Hamiltonians}

We use the \texttt{NuHamil} software~\cite{Miyagi_2023} to generate our Chiral EFT, no-core shell-model Hamiltonian. In particular, we run `NuHamil\_2BME.py' and `NuHamil\_3BME.py' from the \texttt{NuHamil} github repo (\url{https://github.com/Takayuki-Miyagi/NuHamil-public}) to generate the two and three-body terms. In the codes, we set the energy truncation parameters $e_{\max}=8$, $e_{2\max}=16$, and $e_{3\max} =8$ and leave the remaining parameters to be their default values. The generated Hamiltonians are in the $JT$-coupled basis, which we then convert into the $M$-scheme basis, uncoupling the $J$ and $T$ representation so that the nuclear interaction matrix elements are expressed in terms of uncoupled single particle states. %Additionally, we split the data by both $M_2$ and $M_3$ quanta and isospin quantum numbers. For this Hamiltonian, the 1-norm and number of upper diagonal elements for the one-body terms are 6951.46 and 146 respectively.

We evaluate the matrix elements in blocks according to the total two-particle and three-particle $M_J$ (amgular omentum projection) and $M_T$ (isospin projection) quantum numbers, since the Hamiltonian is block-diagonal with respect to these.  The 1-norms and number of non-zero upper diagonal elements in each block are given in Tables \ref{tb:e3max8_2_pp}-\ref{tb:e3max8_3_nnn}.  In these tables, the angular momentum projection for two-body matrix elements is denoted $M_2$ and that for three-body matrix elements by $M_3$.  The isospin projection is translated into its proton-neutron constitution, as noted in the table captions.

\begin{table}[!ht]
\begin{tabular}{|c|c|c|}
\hline
$M_2$       & $\sum_{pqrs}|V_{pqrs}|$ & $|\{V_{pqrs} : |V_{pqrs}| > 0 \: \& \: (pq) \leq (rs)\}|$ \\ \hline
0        & 78180.36  & 133359  \\ \hline
$\pm$ 1  & 72574.36      & 121634  \\ \hline
$\pm$ 2  & 65523.69      & 110715  \\ \hline
$\pm$ 3  & 54796.53  & 90024  \\ \hline
$\pm$ 4  & 43961.42  & 72650  \\ \hline
$\pm$ 5  & 32885.44  & 52542  \\ \hline
$\pm$ 6  & 23509.64  & 37384  \\ \hline
$\pm$ 7  & 15505.49  & 23451  \\ \hline
$\pm$ 8  & 9758.67  & 14499  \\ \hline
$\pm$ 9  & 5600.53  & 7755  \\ \hline
$\pm$ 10  & 3130.95  & 4180  \\ \hline
$\pm$ 11  & 1539.68  & 1818  \\ \hline
$\pm$ 12  & 740.25  & 798  \\ \hline
$\pm$ 13  & 272.54  & 232  \\ \hline
$\pm$ 14  & 101.86  & 71  \\ \hline
$\pm$ 15  & 16.64  & 7  \\ \hline
$\pm$ 16  & 4.18  & 2  \\ \hline
\end{tabular}
\caption{1-norm and number of non-zero upper-diagonal elements of two-body terms per $M_2$-block of the Hamiltonian with $e_{\max}=8$, $e_{2\max}=16$, and $e_{3\max} =8$, restricted to the Proton-Proton (PP, $T_z=-1$) isospin sector.}
\label{tb:e3max8_2_pp}
\end{table}

\begin{table}[!ht]
\begin{tabular}{|c|c|c|}
\hline
$M_2$       & $\sum_{pqrs}|V_{pqrs}|$ & $|\{V_{pqrs} : |V_{pqrs}| > 0 \: \& \: (pq) \leq (rs)\}|$ \\ \hline
0        & 78125.68  & 133359  \\ \hline
$\pm$ 1  & 72623.70      & 121634  \\ \hline
$\pm$ 2  & 65574.73      & 110715  \\ \hline
$\pm$ 3  & 54814.19  & 90024  \\ \hline
$\pm$ 4  & 43979.10  & 72650  \\ \hline
$\pm$ 5  & 32890.32  & 52542  \\ \hline
$\pm$ 6  & 23514.37  & 37384  \\ \hline
$\pm$ 7  & 15506.01  & 23451  \\ \hline
$\pm$ 8  & 9759.10  & 14499  \\ \hline
$\pm$ 9  & 5599.70  & 7755  \\ \hline
$\pm$ 10  & 3130.08  & 4180  \\ \hline
$\pm$ 11  & 1538.85  & 1818  \\ \hline
$\pm$ 12  & 739.41  & 798  \\ \hline
$\pm$ 13  & 272.05  & 232  \\ \hline
$\pm$ 14  & 101.37  & 71  \\ \hline
$\pm$ 15  & 16.47  & 7  \\ \hline
$\pm$ 16  & 4.01  & 2  \\ \hline
\end{tabular}
\caption{1-norm and number of non-zero upper-diagonal elements of two-body terms per $M_2$-block of the Hamiltonian with $e_{\max}=8$, $e_{2\max}=16$, and $e_{3\max} =8$, restricted to the Neutron-Neutron (NN, $T_z=1$) isospin sector.}
\label{tb:e3max8_2_nn}
\end{table}

\begin{table}[!ht]
\begin{tabular}{|c|c|c|}
\hline
$M_2$       & $\sum_{pqrs}|V_{pqrs}|$ & $|\{V_{pqrs} : |V_{pqrs}| > 0 \: \& \: (pq) \leq (rs)\}|$ \\ \hline
0        & 528279.84  & 6049837  \\ \hline
$\pm$ 1  & 496579.62      & 5712741  \\ \hline
$\pm$ 2  & 418333.96      & 4907917  \\ \hline
$\pm$ 3  & 332164.77  & 3851482  \\ \hline
$\pm$ 4  & 248260.81  & 2763218  \\ \hline
$\pm$ 5  & 176369.92  & 1812142  \\ \hline
$\pm$ 6  & 117719.82  & 1079394  \\ \hline
$\pm$ 7  & 74185.50  & 582026  \\ \hline
$\pm$ 8  & 43483.43  & 281665  \\ \hline
$\pm$ 9  & 24020.51  & 123161  \\ \hline
$\pm$ 10  & 12336.62  & 48887  \\ \hline
$\pm$ 11  & 5968.72  & 17876  \\ \hline
$\pm$ 12  & 2609.47  & 5805  \\ \hline
$\pm$ 13  & 1037.00  & 1662  \\ \hline
$\pm$ 14  & 344.51  & 382  \\ \hline
$\pm$ 15  & 95.17  & 69  \\ \hline
$\pm$ 16  & 17.05  & 8  \\ \hline
$\pm$ 17  & 2.09  & 1  \\ \hline
\end{tabular}
\caption{1-norm and number of non-zero upper-diagonal elements of two-body terms per $M_2$-block of the Hamiltonian with $e_{\max}=8$, $e_{2\max}=16$, and $e_{3\max} =8$, restricted to the Proton-Neutron (PN, $T_z=0$) isospin sector.}
\label{tb:e3max8_2_pn}
\end{table}

\renewcommand{\arraystretch}{1.5}

\begin{table}[!ht]
\begin{tabular}{|c|c|c|}
\hline
$M_3$       & $\sum_{pqrstu}|W_{pqrstu}|$ & $|\{W_{pqrstu} : |W_{pqrstu}| > 0 \: \& \: (pq) \leq (rs)\}|$ \\ \hline
$\pm \frac{1}{2}$ & 0.13 & 11 \\ \hline
$\pm \frac{3}{2}$ & 0.12 & 11 \\ \hline
$\pm \frac{5}{2}$ & 0.03 & 3 \\ \hline
$\pm \frac{7}{2}$ & 0.00 & 1 \\ \hline
$\pm \frac{9}{2}$ & 0.00 & 1 \\ \hline
\end{tabular}
\caption{1-norm and number of non-zero upper-diagonal elements of three-body terms per $M_3$-block of the Hamiltonian with $e_{\max}=8$, $e_{2\max}=16$, and $e_{3\max} =8$, restricted to the Proton-Proton-Proton (PPP, $T_z=-\frac{3}{2}$) isospin sector. }
\label{tb:e3max8_3_ppp}
\end{table}

\begin{table}[!ht]
\begin{tabular}{|c|c|c|}
\hline
$M_3$       & $\sum_{pqrstu}|W_{pqrstu}|$ & $|\{W_{pqrstu} : |W_{pqrstu}| > 0 \: \& \: (pq) \leq (rs)\}|$ \\ \hline
$\pm \frac{1}{2}$ & 132.02 & 17707 \\ \hline
$\pm \frac{3}{2}$ & 75.20 & 11823 \\ \hline
$\pm \frac{5}{2}$ & 40.79 & 6681 \\ \hline
$\pm \frac{7}{2}$ & 15.96 & 3006 \\ \hline
$\pm \frac{9}{2}$ & 6.94 & 1248 \\ \hline
$\pm \frac{11}{2}$ & 2.37 & 387 \\ \hline
$\pm \frac{13}{2}$ & 0.83 & 111 \\ \hline
$\pm \frac{15}{2}$ & 0.13 & 16 \\ \hline
$\pm \frac{17}{2}$ & 0.03 & 3 \\ \hline
\end{tabular}
\caption{1-norm and number of non-zero upper-diagonal elements of three-body terms per $M_3$-block of the Hamiltonian with $e_{\max}=8$, $e_{2\max}=16$, and $e_{3\max} =8$, restricted to the Proton-Proton-Neutron (PPN, $T_z=-\frac{1}{2}$) isospin sector. }
\label{tb:e3max8_3_ppn}
\end{table}

\begin{table}[!ht]
\begin{tabular}{|c|c|c|}
\hline
$M_3$       & $\sum_{pqrstu}|W_{pqrstu}|$ & $|\{W_{pqrstu} : |W_{pqrstu}| > 0 \: \& \: (pq) \leq (rs)\}|$ \\ \hline
$\pm \frac{1}{2}$ & 323.17 & 36926 \\ \hline
$\pm \frac{3}{2}$ & 226.01 & 27976 \\ \hline
$\pm \frac{5}{2}$ & 127.61 & 17292 \\ \hline
$\pm \frac{7}{2}$ & 61.31 & 8521 \\ \hline
$\pm \frac{9}{2}$ & 24.54 & 3509 \\ \hline
$\pm \frac{11}{2}$ & 7.48 & 1083 \\ \hline
$\pm \frac{13}{2}$ & 2.34 & 288 \\ \hline
$\pm \frac{15}{2}$ & 0.48 & 42 \\ \hline
$\pm \frac{17}{2}$ & 0.09 & 6 \\ \hline
\end{tabular}
\caption{1-norm and number of non-zero upper-diagonal elements of three-body terms per $M_3$-block of the Hamiltonian with $e_{\max}=8$, $e_{2\max}=16$, and $e_{3\max} =8$, restricted to the Proton-Neutron-Neutron (PNN, $T_z=\frac{1}{2}$) isospin sector. }
\label{tb:e3max8_3_pnn}
\end{table}

\begin{table}[!ht]
\begin{tabular}{|c|c|c|}
\hline
$M_3$       & $\sum_{pqrstu}|W_{pqrstu}|$ & $|\{W_{pqrstu} : |W_{pqrstu}| > 0 \: \& \: (pq) \leq (rs)\}|$ \\ \hline
$\pm \frac{1}{2}$ & 0.13 & 11 \\ \hline
$\pm \frac{3}{2}$ & 0.12 & 11 \\ \hline
$\pm \frac{5}{2}$ & 0.02 & 3 \\ \hline
$\pm \frac{7}{2}$ & 0.00 & 1 \\ \hline
$\pm \frac{9}{2}$ & 0.00 & 1 \\ \hline
\end{tabular}
\caption{1-norm and number of non-zero upper-diagonal elements of three-body terms per $M_3$-block of the Hamiltonian with $e_{\max}=8$, $e_{2\max}=16$, and $e_{3\max} =8$, restricted to the Neutron-Neutron-Neutron (NNN, $T_z=\frac{3}{2}$) isospin sector. }
\label{tb:e3max8_3_nnn}
\end{table}

\section{Algorithm for block decomposition of chiral EFT Hamiltonians}\label{app:chiral_block_decomp}
Here we outline an algorithm for implementing no-shell core-model Hamiltonians with up to three-body terms that exploits the symmetry of the Hamiltonian with respect to the total two-particle and three-particle $M_J$ (angular momentum projection) and $M_T$ (isospin projection) quantum numbers. Since the Hamiltonian is block diagonal with respect to these quantum numbers, we can employ a similar strategy as was applied to the phenomenological Hamiltonians in the main text to separate the simulation into blocks that can be implemented separately. However, for the no-core shell-model Hamiltonians, such a strategy results in worse resource estimates than can be obtained by varying the space/time trade off parameters in the data loaders for the state preparation subroutines. Nevertheless, for completeness we outline the strategy and present the resource estimates here.

In the $M$ scheme, the two and three-body terms are separately block-diagonalized by different $M$ quantum numbers, which we call $M_2$ and $M_3$. Loosely speaking, $M_2$ comes from summing up the angular momentum projection of pairs of single-particle basis states, as in the case of the shell-model Hamiltonians, and analogously, $M_3$ comes from summing up the angular momentum projection of triplets of single-particle basis states.

We can construct an algorithm that takes advantage of this block structure, akin to the one for the shell-model Hamiltonian: For a given Hamiltonian, we compute the number of $M_3$ values, which we store in a list $\mathbf{M}_3$. For each $M_3\in \mathbf{M}_3$, we compute the number of $M_2$ values that are involved in the considered $M_3$-block. More specifically, let the $j$ values in the number of levels in a given valence space range from $\frac{1}{2}$ to $j_{\max}$. Then, for each $M_3$ value, the allowed $M_2$ values will be given by
\begin{equation}
    \max\{ -j_{\max}, M_3 -j_{\max}\} \leq M_2 \leq \min\{j_{\max}, M_3 +j_{\max} \},
\end{equation}
which we store in a list $\mathbf{M}_2$.~\footnote{One way to understand this is as follows. Without loss of generality, consider an example where $j_{\max} = \frac{7}{2}$. A three-body state in this valence space can have a maximum total $M$ value of $M_3 = \frac{21}{2}$. The extremal $|M_3|$-values are (i) $\frac{21}{2}$ and (ii) $\frac{1}{2}$. For $M_3=\frac{21}{2}$, we can only have $M_2 = 7$ as we have to combine this with a single $m=\frac{7}{2}$ state. For $M_3=\frac{1}{2}$, the maximum $M_2$ value is 4 as this can be combined with a single $m=\frac{7}{2}$ state, whereas the minimum $M_2$ value is -3 as this can be combined with a single $m=\frac{7}{2}$ state.} Then, for each $M_3\in \mathbf{M_3}$, we perform a phase estimation on a Hamiltonian that is the sum of the one-body term, the chosen $M_3$-block, and all the $M_2$-blocks for all $M_2 \in \mathbf{M}_2$; we store the phase estimation outcomes in a list $\mathbf{E}$, and aggregate the gate counts over all $M_3$ values. In the context of ground-state energy estimation, for example, the lowest estimate in $\mathbf{E}$ will be the lowest estimate of the Hamiltonian's energy, given an input state. As in the phenomenological shell-model Hamiltonian, the overall algorithm a number ($|\mathbf{M_3}|$) of phase estimation runs, and the number of logical qubits and physical size of a quantum computer needed to run this algorithm will be determined by the largest coherent phase estimation circuit instance.

A similar argument can be followed to decompose the Hamiltonian into blocks with the same isospin projection quantum number, since this is also a symmetry of the Hamiltonian. As the isospin projection operator commutes with the total angular momentum operator, we can also apply these block decompositions independently to separately simulate blocks with both a given total angular momentum projection and isospin projection.

\begin{figure}[ht!]
    \centering
    \includegraphics[width=0.65\linewidth]{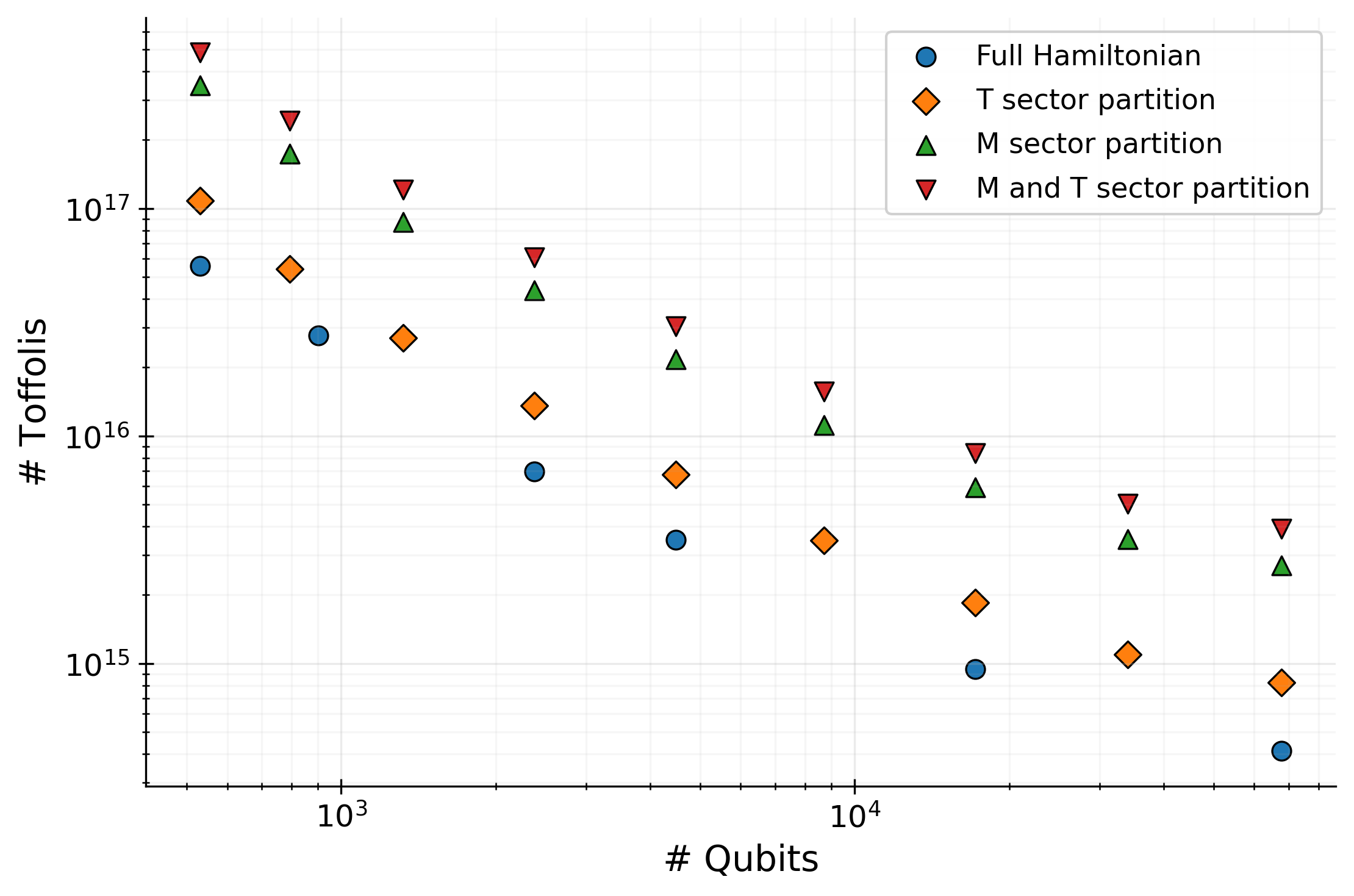}
    \caption{Quantum resource estimates for applying quantum phase estimation to the no-core shell-model Hamiltonian explored in Sec.~\ref{sec:qre}. Each set of data corresponds to a partitioning strategy, with the blue circles corresponding to the full Hamiltonian with no partitioning into symmetry sectors, the orange diamonds corresponding to a partitioning into sectors with different isospin quantum numbers, the green triangles corresponding to partitioning into sectors with different total angular momentum projection quantum numbers and the red inverted triangles corresponding to a partitioning by both isospin projection and total angular momentum quantum numbers. The variation in Toffoli count/number of qubits within each set of data corresponds to a Pareto-optimal search over QROM space/time tradeoff parameters.}
    \label{fig:qre-analysis-sector-partitions}
\end{figure}

As shown in Fig.~\ref{fig:qre-analysis-sector-partitions}, the optimal strategy for this Hamiltonian is \textit{not} to simulate different sectors separately at all, but rather to perform phase estimation on the Hamiltonian as a whole. The reason for this is that the partitioning into different sectors always involves some redundant terms being included (for example, the entirety of the one-body terms in the phenomenological models are included for every $M_2$ sector). For the phenomenological Hamiltonians, this was acceptable, since the dominant terms were the two-body interactions, which contained no such redundancy.

By contrast, for the no-core shell-model Hamiltonians, we are required to duplicate some of the two-body terms by including them in simulations corresponding to multiple $M_3$ sectors. Since the two-body interactions are still the dominant terms in the Hamiltonian by 1-norm and number of non-zero upper diagonal matrix elements, this increases the cost of the strategy substantially, rendering it ineffective for this Hamiltonian.

\nocite{}

\end{document}